\documentclass[draft]{agujournal2019}
\usepackage{url} 
\usepackage{lineno}
\usepackage[inline]{trackchanges} 
\usepackage{soul}
\usepackage{algorithm}
\usepackage{algpseudocode}
\usepackage{subcaption}
\usepackage{bbm}
\usepackage{amsmath}

\DeclareMathOperator*{\argmin}{argmin}
\usepackage{amssymb}
\usepackage{tikz}
\usetikzlibrary{positioning} 
\usetikzlibrary{calc}
\draftfalse

\journalname{Journal of Advances in Modeling Earth Systems}

\begin{document}

%
%


\title{Entropic learning enables skilful forecasts of ENSO phase at up to two years lead time}

%
%




\authors{Michael Groom\affil{1}, Davide Bassetti\affil{2}, Illia Horenko\affil{2}, Terence J. O'Kane\affil{3}}


\affiliation{1}{CSIRO Environment, Eveleigh, New South Wales, Australia}
\affiliation{2}{Faculty of Mathematics, Rheinland-Pfälzische Technische Universität Kaiserslautern Landau, Kaiserslautern, Germany}
\affiliation{3}{CSIRO Environment, Hobart, Tasmania, Australia}




\correspondingauthor{Michael Groom}{Michael.Groom@csiro.au}



\begin{keypoints}
\item A novel approach based on ensembles of entropic learning models is shown to perform skilful forecasts of ENSO phase at up to 24 months lead.
\item Our approach effectively mitigates overfitting and delivers probabilistic forecasts with skill comparable to the IRI ENSO prediction plume.
\item Successful hindcast validation of major ENSO events confirms the method's operational potential for interannual climate prediction.
\end{keypoints}

%
%

%
%


\begin{abstract}
This paper extends previous work (Groom et al., \emph{Artif. Intell. Earth Syst.}, 2024) in applying the entropy-optimal Sparse Probabilistic Approximation (eSPA) algorithm to predict ENSO phase, defined by thresholding the Ni\~no3.4 index. Only satellite-era observational datasets are used for training and validation, while retrospective forecasts from 2012 to 2022 are used to assess out-of-sample skill at lead times up to 24 months. Rather than train a single eSPA model per lead, we introduce an ensemble approach in which multiple eSPA models are aggregated via a novel meta-learning strategy. The features used include the leading principal components from a delay-embedded EOF analysis of global sea surface temperature, vertical temperature gradient (a thermocline proxy), and tropical Pacific wind stresses. Crucially, the data is processed to prevent any form of information leakage from the future, ensuring realistic real-time forecasting conditions. Despite the limited number of training instances, eSPA avoids overfitting and produces probabilistic forecasts with skill comparable to the International Research Institute for Climate and Society (IRI) ENSO prediction plume. Beyond the IRI’s lead times, eSPA maintains skill out to 22 months for the ranked probability skill score and 24 months for accuracy and area under the ROC curve, all at a fraction of the computational cost of a fully-coupled dynamical model. Furthermore, eSPA successfully forecasts the 2015/16 and 2018/19 El Ni\~no events at 24 months lead, the 2016/17, 2017/18 and 2020/21 La Ni\~na events at 24 months lead and the 2021/22 and 2022/23 La Ni\~na events at 12 and 8 months lead.
\end{abstract}

\section*{Plain Language Summary}
This study introduces a new, cost-effective way to forecast the phase of the El Ni\~no–Southern Oscillation (ENSO) -- the dominant mode of interannual climate variability in the Pacific that alternates between El Ni\~no, La Ni\~na, and neutral phases -- up to two years in advance using a novel machine learning method called the entropy-optimal Sparse Probabilistic Approximation (eSPA) algorithm. Despite relying solely on observational and assimilated data from the satellite era (circa 1980 onwards), eSPA overcomes the common problem of having too few historical events to learn from, as it is designed to avoid overfitting to noise in high-dimensional data. The method delivers forecasts with skill comparable to those produced by the well-established International Research Institute for Climate and Society (IRI) ENSO prediction plume, while requiring far less computing power to generate its predictions. In summary, this work demonstrates that advanced machine learning techniques can improve long-range ENSO forecasts, offering a promising tool for better preparing for the broad societal and economic impacts associated with global climate variability.

\section{Introduction} \label{sec:intro}

Seasonal-to-interannual forecasts of the El Ni\~no–Southern Oscillation (ENSO) are of great practical importance due to its far-reaching impacts on global weather patterns, ecosystems, and economies. However, predicting ENSO events more than 12 months in advance remains extremely challenging. Both physics-based dynamical models and statistical approaches tend to lose skill at longer lead times, especially when forecasts must cross the boreal spring predictability barrier -- a well-known limitation that persists even in state-of-the-art coupled models \cite{OKane2020}. Moreover, the short observational record (e.g. the satellite era from circa 1980 onward only contains a few strong ENSO events) means that purely data-driven models face an acute small data problem \cite{Horenko2020}, with only a limited number of instances of high-dimensional data to train on. These factors have contributed to the general lack of accuracy in long-range ENSO forecasts despite decades of research. One approach to improving skill is the use of multi-model ensembles (MMEs), which tend to have higher skill than predictions from a single model \cite{Tippett2008}, and which provide a straightforward approach to quantifying forecast uncertainty due to uncertainty in model formulation \cite{Kirtman2014}. The International Research Institute for Climate and Society (IRI) ENSO prediction plume exemplifies this approach by aggregating forecasts from the world's leading climate prediction centres, making it the operational benchmark for ENSO forecasting. An up-to-date version of the plume can be found at the IRI website (\url{https://iri.columbia.edu/our-expertise/climate/forecasts/enso/current/?enso_tab=enso-sst_table}).

Since its introduction in 2002, the IRI plume has been continuously improved and updated through the addition of new models as well as the application of systematic bias corrections and ensemble calibrations to increase reliability. 
Traditionally, dynamical models have held a slight edge over statistical models for ENSO prediction on seasonal timescales. An assessment by \citeA{Barnston2012} on the models comprising the IRI plume throughout 2002-2011 found that dynamical models produced slightly more accurate forecasts through the boreal spring, although overall skill was low for all methods beyond about 6–9 months. Statistical models were also shown to suffer from slippage to a greater degree, which is the tendency for predicted transitions to lag observed transitions in the ENSO state due to a bias toward persistence. \citeA{Tippett2012} conducted a probabilistic skill assessment of the IRI plume over the same period, using the entire MME to compute probabilities for each phase of ENSO (i.e., El Ni\~no, La Ni\~na or neutral). Forecasts at longer lead times failed to capture the initiation and termination of events and exhibited the same slippage problem as the deterministic forecasts. Statistical post-processing, in the form of a multiple linear regression, was shown to generally be effective in removing slippage. 

\citeA{Barnston2015} showed that removing each forecast model’s mean bias (and amplitude bias where necessary) before combining the models to form the MME improved short-lead forecasts and produced a more representative ensemble spread. Following the findings of \citeA{Tippett2008}, individual ensemble members of all models were weighted equally when combining them to form the MME mean forecast as apparent skill differences between models tend to be indistinguishable from sampling error over typical hindcast periods of 20-30 years. This results in models with larger ensembles being weighted more heavily in the MME mean. \citeA{Barnston2015} also showed that historical hindcast skill should be used to determine forecast uncertainty rather than the models' ensemble spreads, as these produce a less reliable distribution (in the sense of predicted probabilities of events being well calibrated with observed frequencies of those events). Other methods for improving reliability include calibrations derived from regressing past model outputs onto observations. \citeA{Tippett2014} showed that care must be taken when doing this, as sampling error results in the regression-corrected probability forecasts being systematically overconfident. Estimating the regression parameters using shrinkage methods such as ridge regression substantially reduces this overconfidence. 

In tandem with post-processing refinements, significant gains have also come from incorporating improved dynamical models, of which the most impactful are the coupled models comprising the North American Multimodel Ensemble (NMME) project \cite{Kirtman2014} which began producing real-time forecasts in August 2011. \citeA{Barnston2019} evaluated the deterministic skill of ENSO hindcasts made by the NMME against that of real-time forecasts by the IRI plume over 2002-2011. The top two performing individual models from the NMME were found to be the NOAA/NCEP CFSv2 and the Canadian CMC2 models. The NMME was also shown to have a slightly higher anomaly correlation skill for the shortest lead times, with this difference increasing with increasing lead times. Similar results were observed when comparing NMME model hindcasts over 1981-2010 with available hindcasts over the same period from IRI models. \citeA{Tippett2019} conducted a probabilistic skill assessment of the NMME over 1982-2015 (containing both hindcasts and real-time forecasts) and computed the ranked probability skill score (RPSS) and the logarithmic skill score (LSS) for probabilistic forecasts of three, five and seven categories defining the phase of ENSO (determined by varying the number of thresholds of the Ni\~no3.4 index). Comparisons of the three-category RPSS against the earlier results for the IRI plume presented in \citeA{Tippett2012} demonstrate that skill is most improved for target months from June to August at lead times of 0-3 months, along with October to March at lead times greater than 7 months. An important caveat to this comparison is that from 2002-2011 the IRI used a different definition for ENSO phase, defining El Ni\~no and La Ni\~na events as anomalies in the Ni\~no3.4 region that fall in the upper/lower quartile of the climatological distribution for a given season.

In recent years, the advent of deep learning has sparked a resurgence of interest in data-driven ENSO forecasting. A prominent example is the work of \citeA{Ham2019}, who trained a convolutional neural network (CNN) using a transfer-learning approach – first on large collections of climate model simulations from the CMIP5 ensemble, and then on ocean reanalysis data – to predict the Ni\~no3.4 index $n$ months ahead based on sea surface temperatures and oceanic heat content from the current and previous 2 months. This deep learning model outperformed state-of-the-art dynamical models at lead times beyond 6 months, achieving a pattern correlation with the observed index above 0.5 out to 17 months. A follow-up study applied a multitask learning framework to further improve forecast accuracy by addressing the seasonally varying nature of ENSO precursors \cite{Ham2021}. Other researchers have explored more advanced architectures and regularisation strategies to push predictive skill to even longer lead times. For instance, forecasts generated by the 3D transformer model of \citeA{Zhou2023} were found to be skilful in predicting the Ni\~no3.4 index at up to 18 months lead time, although biases in the training data (coming from biases in the underlying CMIP6 climate models generating the data) led to reduced skill in other regions of the Pacific. A few studies have also attempted long-range ENSO prediction using only observational and reanalysis data. Notably, \citeA{Patil2023} developed a deep CNN model trained on observed/reanalysed sea surface and vertically-averaged subsurface temperatures, with skilful forecasts obtained out to 20 months lead time. Their CNN model featured multiple forms of regularisation including dropout, as well as average pooling to reduce the number of model parameters. Similar to \citeA{Ham2021}, it also contained heterogeneous parameters for each target season to account for seasonal variations in precursors, establishing it as a prime example of the state-of-the-art performance that is obtainable with deep learning for long-range ENSO prediction. In March 2025 this model was added to the IRI plume.

While these results demonstrate the promise of modern machine learning for multi-year ENSO forecasting, they also highlight persistent challenges. Many deep learning methods require "big data", currently only obtainable through large climate model ensembles, to train models with enough parameters to capture complex spatiotemporal patterns, which can result in them inheriting some of the biases in the training data. In contrast, methods trained solely on the limited observational record risk overfitting unless they are specifically designed for the "small data" regime. To address these challenges, recent work has proposed entropic learning techniques which are based on sparsified, probabilistic approximations of the data that employ the principle of maximum entropy from information theory to avoid overfitting to noisy/uninformative features \cite{Horenko2020,Horenko2022,Vecchi2022,Horenko2023,Vecchi2024}. A comparative study by \citeA{Groom2024} applied the entropy-optimal Sparse Probabilistic Approximation (eSPA) classifier to long-range prediction of ENSO phase and found that it can match or exceed the accuracy of deep neural networks while requiring orders of magnitude less training time and number of parameters. Building on that foundation, the present study focuses on ENSO phase forecasting using only real-time observational and reanalysis data from the satellite era, without any reliance on climate model simulations. A suite of hindcast experiments covering 2012–2022, with lead times up to 24 months, are performed to rigorously evaluate out-of-sample forecast skill, with the combined (model-based) probabilistic forecasts produced from the IRI plume over the same period employed as a benchmark. While technically hindcasts, great care is taken to ensure real-time conditions are enforced to make the comparison with the IRI plume as valid as possible. The period of 2012-2022 is chosen since (i) NMME models such as CFSv2 were introduced in the IRI plume starting from mid-2011 and (ii) in January 2012 the definition of ENSO phase was switched from the earlier definition based on quartiles to use a $\pm0.45^\circ$ threshold of the Ni\~no3.4 index, which was updated to a $\pm0.5^\circ$ threshold in May 2013. 

The remainder of this paper is organised as follows. Section \ref{sec:methods} describes the datasets, pre-processing, and the entropic learning methodology used for ENSO phase forecasting. Section \ref{sec:results} presents the forecasting results and comparisons with the IRI plume, including skill assessments stratified by lead time and target season. Section \ref{sec:conclusion} discusses the implications of the findings and concludes the paper.

\section{Materials and Methods}  \label{sec:methods}

\subsection{Datasets}
In this study, only observations and reanalyses from the satellite era are employed when training and validating the entropic learning models. Unlike the earlier study of \citeA{Groom2024}, both oceanic and atmospheric fields are considered to give a more complete picture of the coupled dynamics of ENSO. The oceanic fields considered are monthly means of global sea surface temperature (SST) between $60^\circ$S-$60^\circ$N and the vertical derivative of subsurface temperature ($\mathrm dT/\mathrm dz$) between $40^\circ$S-$40^\circ$N and restricted to longitudes of $120^\circ$E-$80^\circ$W and depths of $0$-$700$m. The atmospheric fields considered are monthly means of the zonal and meridional surface wind stresses ($\tau_x$ and $\tau_y$), restricted to latitudes of $20^\circ$S-$20^\circ$N and longitudes of $120^\circ$E-$80^\circ$W, with a mask is applied so that only oceanic wind stresses are selected.

The SST data are taken from the NOAA Optimum Interpolation Sea Surface Temperature (OISST) V2.1 product \cite{Huang2021} and are provided on a $0.25^\circ\times0.25^\circ$ global grid. The $\mathrm dT/\mathrm dz$ data are derived from potential temperature fields taken from the NOAA/NCEP Global Ocean Data Assimilation System (GODAS) reanalysis \cite{Behringer2004} and are on a $1^\circ\times1/3^\circ$ grid with 40 vertical levels. Note that SST in GODAS is strongly nudged towards the weekly OISST data with a relaxation time of 5 days \cite{Xue2012}. The wind stress data are taken from the NCEP/DOE Reanalysis 2 (NNR2) dataset \cite{Kanamitsu2002}, which provides the momentum flux, heat flux and freshwater flux forcings to GODAS, and are provided on a $2.5^\circ\times2.5^\circ$ global grid. The combined data range from September 1981 to present day (currently December 2024), giving a total of 520 monthly averages. The start date of the first hindcast in January 2012 also ensures that there are at least 30 years of training data used to define anomalies and calculate EOFs as described below.

\subsection{Pre-processing} \label{subsec:pre-processing}
To calculate the Ni\~no3.4 index, a 30-year sliding climatology is used when calculating the anomalies in the Ni\~no3.4 region for a given year. This ensures that no information from the future leaks into a given hindcast. Rather than recalculate all of the previous anomalies each time the climatological base period is updated, as is common in operational settings, in this study they are kept fixed once first calculated. This ensures that the index, and thus the class labels, are uniquely defined across the entire period using only data that was available at the time. It also acts as a mild form of detrending, since each anomaly appearing earlier in the dataset is with respect to a local base period rather than a fixed global base period that is typically defined in the most recent part of the dataset and thus in a warmer climate. To enable anomalies in the first 30 years of the dataset to be calculated in this manner, the sliding climatology needs to be defined by augmenting with SST data in the Ni\~no3.4 region from earlier than September 1981. This data is taken from the NOAA Extended Reconstructed SST V5 (ERSSTv5) dataset \cite{Huang2017}.

To produce the SST and $\mathrm dT/\mathrm dz$ fields, the OISST and GODAS data are first regridded to a $1^\circ\times1^\circ$ global grid using a conservative remapping and ensuring that a common land-sea mask is used. Note that while this step is not strictly necessary for this study since an EOF analysis is subsequently performed on each field separately, it allows for the possibility of using the full fields as direct inputs that are defined on a common grid in future studies. Following this, the vertical derivatives are calculated for the GODAS data. To generate the features used in each hindcast the following steps are performed, which ensure there is no leakage of future information into a given hindcast:
\begin{enumerate}
    \item The seasonal cycle is removed by converting the data to anomalies. A base period of January 1982 to December of the year prior to the start date of the hindcast is used to define the climatology.
    \item A linear detrending step is performed prior to the EOF calculation to remove the global warming signal, where the trend is first calculated over the same base period as the anomalies and then extrapolated to times outside of this period.
    \item An Empirical Orthogonal Function (EOF) analysis is performed as a dimensionality reduction step. To preserve the validity of the Euclidean distance metric, which is used both to define the reconstruction error of the principal component decomposition in EOF analysis as well as the discretisation error in eSPA, we employ the \citeA{Takens1981} delay embedding theorem and embed $n$ lags of the data prior to constructing the covariance matrix.\footnote{According to Takens' theorem, the mapping of an attractor with box-counting dimension $d$ into the $k$-dimensional embedded space is diffeomorphic when $k>2d$. In practice, it can be difficult to estimate $d$ and therefore the embedding length $n$ is chosen empirically by testing a range of different values. The results given in section \ref{sec:results} use an embedding length of 12 months, which was found to give good results while also being consistent with embedding lengths used in other studies on ENSO prediction \cite{Zhou2023} as well as capturing known optimal growth times of SST anomalies in the Pacific \cite{Lou2021}.} Using a slight abuse of terminology, we refer to this procedure as Singular Spectrum Analysis (SSA) to distinguish it from conventional EOF analysis.
    \item The SSA modes are calculated for the same base period of January 1982 to December of the year prior to the start date of the hindcast, then projected onto the (lagged) anomalies to give the full time series of principal components (PCs).
\end{enumerate}
This procedure is then repeated for each year of hindcasts from $2012,\ldots,2022$. Note that step 1 introduces an inconsistency between the climatology used to define the index and that used to define the SST, $\mathrm dT/\mathrm dz$ and wind stress anomalies, which are re-calculated each year. While it would be possible to employ a similar 30-year sliding climatology for the SST and wind stress datasets (for example by augmenting them with data prior to 1981 from the ERSSTv5 and NNR1 \cite{Kalnay1996} datasets), a lack of high-quality subsurface ocean data prior to 1980 prevents us from doing this for GODAS as well. Instead, the decision was made to remove the seasonal cycle over the same period that the EOFs are calculated for in step 3, with the linear detrending in step 2 acting to reduce the inconsistency in climatologies. 

Following \citeA{Groom2024}, a fixed percentage of the total variance is used to select the number of principal components that are retained as features. For global SST, 160 PCs explaining $\sim80\%$ of the total variance are retained, for $\mathrm dT/\mathrm dz$ 140 PCs are retained explaining $\sim80\%$ of the total variance while for the wind stresses 100 PCs are retained explaining $\sim70\%$ of the total variance. To improve skill at short lead times, the monthly Ni\~no3.4 index is added as a feature along with the warm water volume index -- defined as the anomalous integrated depth of the $20^\circ$ isotherm (Z20) over the domain $120^\circ$E-$80^\circ$W and $5^\circ$S-$5^\circ$N \cite{Meinen2000} -- bringing the total number of (real-valued) features used by the model to 402. Upon assembly of the feature matrix, a final pre-processing step of mapping the data to a uniform distribution with values between 0 and 1 using a quantile transformation is applied to all of the features. This is performed separately for each hindcast, which only contains data up until its given start date, thus avoiding any leakage of future information when calculating the empirical cumulative distribution function for each feature.

\begin{figure}[t]
\centering
\noindent\includegraphics[width=\textwidth]{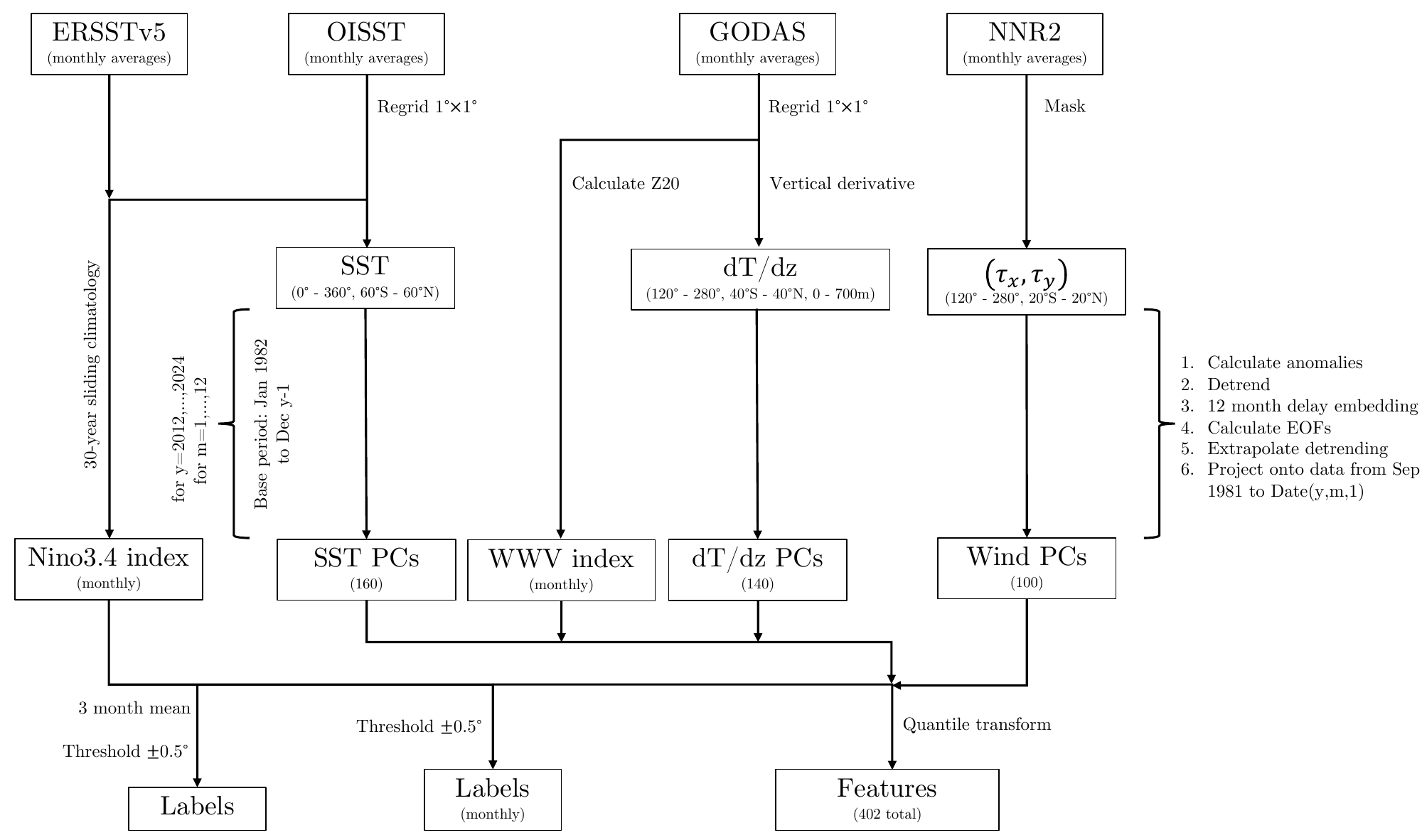}
\caption{A summary of the data pre-processing steps described in section \ref{subsec:pre-processing}.}
\label{fig:data-summary}
\end{figure}

The targets for prediction (class probability distributions) are generated by considering the probability of the Ni\~no3.4 index being greater than $0.5^\circ$C (El Ni\~{n}o), less than $-0.5^\circ$C (La Ni\~{n}a) or neither (neutral) in $n$ months time. For consistency with the IRI ENSO prediction plume, the 3-month running average of the Ni\~no3.4 index is used, which along with the threshold of $\pm0.5^\circ$C gives class proportions of 0.25, 0.46 and 0.29 for the La Ni\~{n}a, neutral and El Ni\~{n}o classes over the period of September 1981 to December 2024. These are labelled as classes 1, 2 and 3 respectively when calculating metrics that depend on the ordinal ranking of classes such as the ranked probability score. No correction is made for the slight imbalance of classes. Also note that from January 2012 to April 2013 the definition of the classes is inconsistent with that used by the IRI plume (which used a threshold of $\pm0.45^\circ$ over this period). This is expected to produce only minor differences in the evaluation of its skill. A summary of the data pre-processing methodology employed for generating forecasts is given in Figure \ref{fig:data-summary}.

\begin{figure}[H]
  \centering
  \begin{tikzpicture}
    \node (img1) {\includegraphics[width=\textwidth]{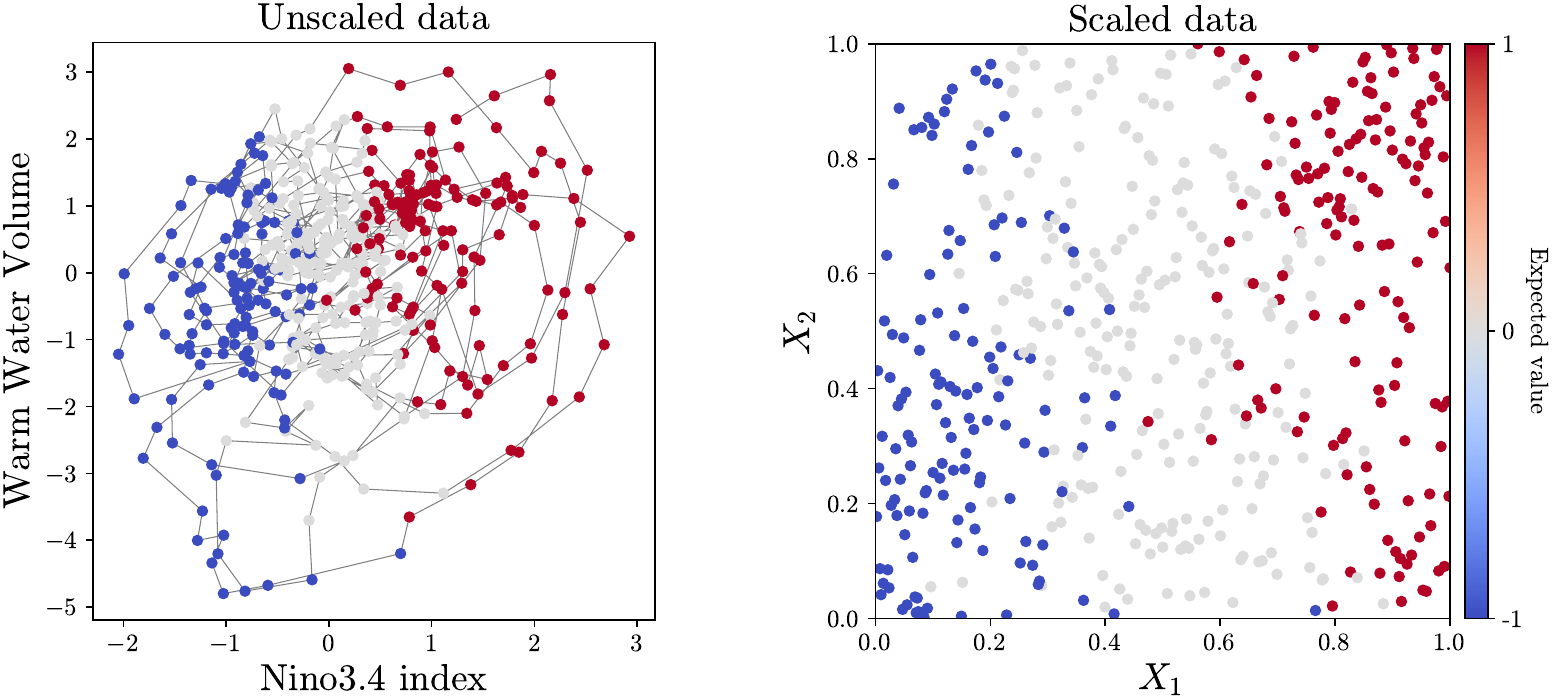}};
    \node (img2) [below=0.75cm of img1, draw, thick, rounded corners] {\includegraphics[width=\textwidth]{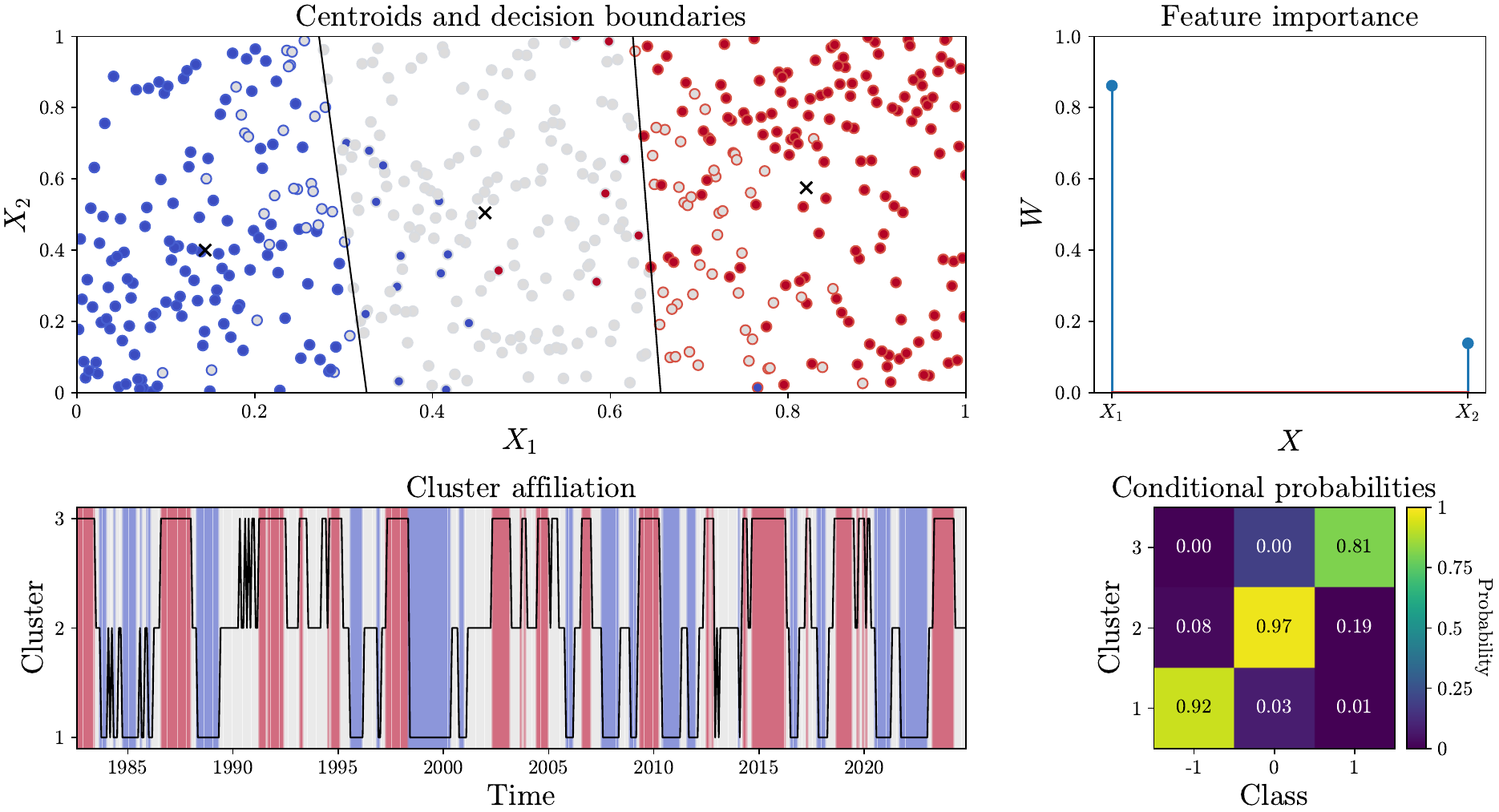}};
    \coordinate (start) at ($ (img1.south west)!0.745!(img1.south east) $);
    \coordinate (end) at ($ (img2.north west)!0.745!(img2.north east) $);
    \draw[->, black, line width=1.5pt] (start) -- (end);
    \draw[->, black, line width=1.5pt] 
      ($ (img1.west)!0.435!(img1.east) $) -- 
      ($ (img1.west)!0.495!(img1.east) $);
    \node[anchor=north west, xshift=0pt, yshift=0pt] at (img1.north west) {(a)};
    \coordinate (b) at ($ (img1.north west)!0.48!(img1.north east) $);
    \node[anchor=north west, xshift=0pt, yshift=0pt] at (b) {(b)};
    \node[anchor=north west, xshift=0pt, yshift=0pt] at (img2.north west) {(c)};
    \coordinate (d) at ($ (img2.north west)!0.655!(img2.north east) $);
    \node[anchor=north west, xshift=0pt, yshift=0pt] at (d) {(d)};
    \coordinate (e) at ($ (img2.north west)!0.54!(img2.south west) $);
    \node[anchor=north west, xshift=0pt, yshift=0pt] at (e) {(e)};
    \coordinate (topPoint) at ($ (img2.north west)!0.655!(img2.north east) $);
    \coordinate (bottomPoint) at ($ (img2.south west)!0.655!(img2.south east) $);
    \coordinate (f) at ($ (topPoint)!0.54!(bottomPoint) $);
    \node[anchor=north west, xshift=0pt, yshift=0pt] at (f) {(f)};
  \end{tikzpicture}
  \caption{An illustration of the data processing and entropic learning procedure for a simplified version of the full learning problem. (a) The data, consisting of the Ni\~no3.4 and WWV indices, are plotted in phase space, coloured by the 1-month ahead class labels. (b) A quantile transformation is performed to map each feature to be uniformly distributed on the interval $[0,1]$. An eSPA model with 3 clusters is then fitted to this data. (c) The cluster centroids (given by $C$) and decision boundaries are plotted in the transformed space where each dimension is scaled by $\sqrt{W_d}$. The predictions for each data point (in terms of expected value) are given as the edge colour for each marker. (d) The feature importance $W_d$ for both dimensions of the dataset. (e) The cluster that each data point is assigned to as a function of time (given by $\Gamma$). The background shading corresponds to the true class 1-month ahead of time $t$ (given by $\Pi$). (f) The predicted probabilities (given by $\Lambda$) for each class $m$, conditioned on an instance being assigned to cluster $k$. For further details on the structure of an eSPA model, please see \ref{app:eSPA}.}\label{fig:eSPA-summary}
\end{figure}
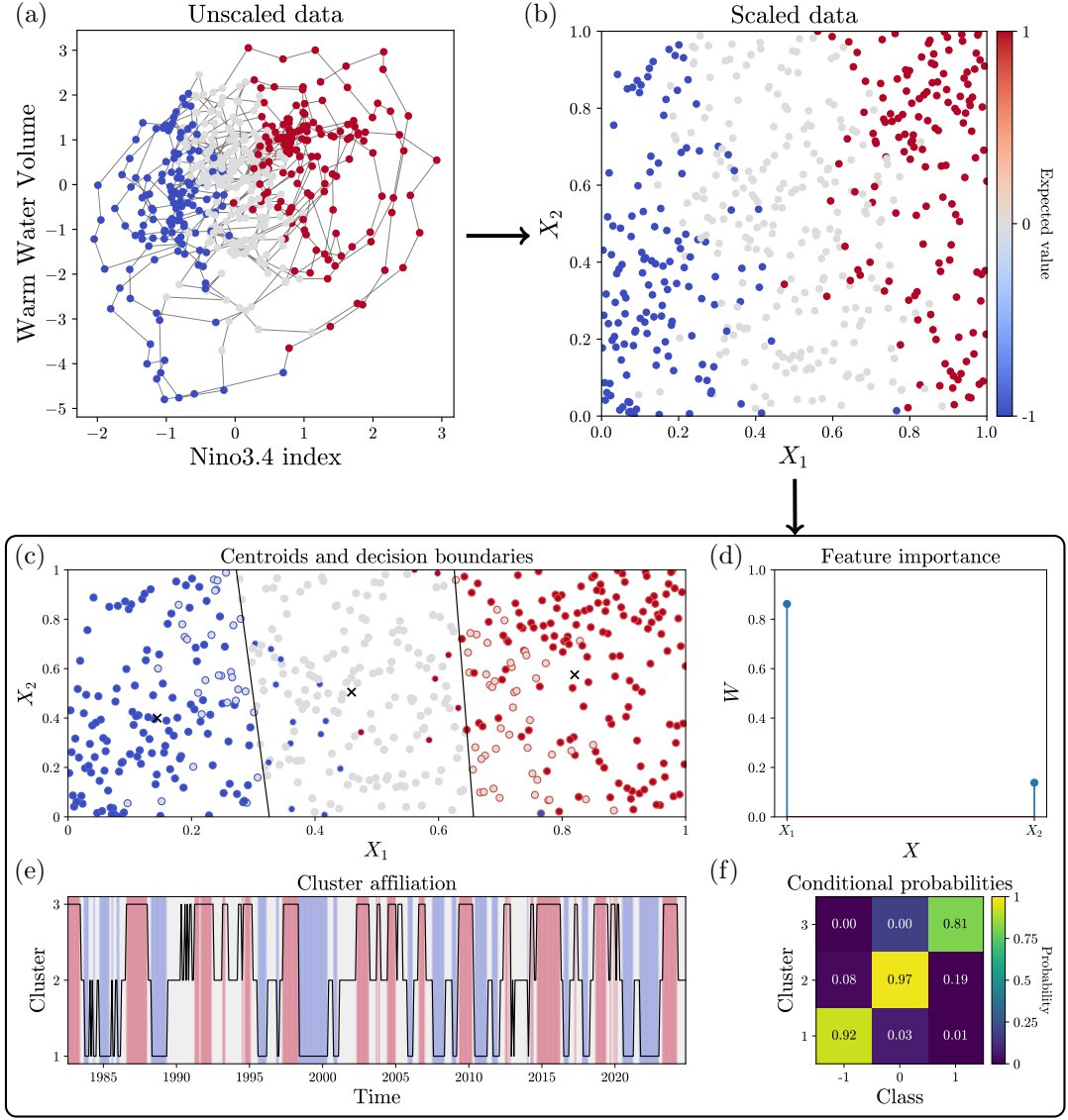

\subsection{Entropic learning}

Given the desire to use only observations and reanalyses from the satellite era, the limited number of instances available for learning (ranging from 350 for the earliest hindcast to 520 as of December 2024) relative to the number of features (406) makes the prediction task a supervised learning problem in the small data regime where the risk of overfitting, for a given model complexity, is far greater than in typical big data applications. The recently proposed eSPA classifier has been demonstrated to cheaply and effectively avoid overfitting in this regime \cite{Horenko2020,Vecchi2022} and has been thoroughly assessed on the problem of ENSO phase prediction in \citeA{Groom2024}. \ref{app:eSPA} gives an overview of the eSPA algorithm. To aid in understanding, a visual depiction of the components comprising a fitted eSPA model is given in Figure \ref{fig:eSPA-summary} for a simplified version of the problem that uses just the Ni\~no3.4 index and warm water volume (i.e. two variables commonly used to define the phase of the ENSO recharge/discharge oscillator \cite{Timmermann2018}) as features for a 1-month lead time prediction.

Compared to the simple out-of-sample prediction problems used in \citeA{Vecchi2022,Horenko2023,Vecchi2024,Groom2024}, the formulation of the problem in this study is targeted towards the generation of real-time forecasts. This presents several additional impediments, many of which are due to the non-stationarity of the dynamical system we are trying to predict, that all act to reduce skill at longer lead times relative to the ideal case. Firstly, due to the inability to label instances for lead times with target dates beyond the start date of the forecast, there is an increasingly larger gap between the end of the training set and the start date of the forecast as lead time increases. The result of this is that the end of the training set becomes increasingly less relevant to the current conditions from which we are trying to generate the forecast, which is referred to as concept drift in the machine learning literature \cite{Gama2014}. Secondly, the predictions for each lead time are all made from a single instance, i.e. the latest available monthly-averaged data. This necessitates some form of model selection, since a given model may make predictions that are otherwise correct but are incorrect for that particular instance.

We attempt to mitigate both of these issues by using an ensemble of models to generate individual predictions for each lead time and then aggregate these predictions to give a final prediction for that lead time. A more advanced aggregation strategy that leverages the interpretability of eSPA is described below in section \ref{subec:post-processing}, but prior to this an arithmetic average is used. One option for generating the ensemble is to fit eSPA models using all of the available data with different initial guesses for the model parameters, since each initial guess is guaranteed to converge to a local minimum of the loss function that will, in general, be different for different initial guesses. However, in practice we find that it is better to first split the data into a training and validation set and then fit multiple eSPA models on the training data as this also allows for hyperparameter tuning to be performed. The validation set is used to select the best model, according to a particular metric (see section \ref{subsec:metrics} for details), across all initial guesses and hyperparameter combinations and then this process is repeated for different splits of the data until a sufficient ensemble size is generated. A total of 50 such cross-validation splits are employed, each of which is stratified by class so that the proportions of El Ni\~{n}o, La Ni\~{n}a and neutral events are the same for both training and validation sets across all splits. A train/validation split of 80\%/20\% was found to provide a good trade-off between a large enough training set to avoid issues of non-stationarity when training a given model and a large enough validation set to accurately assess its generalisation to unseen data. 

As in \citeA{Groom2024}, we train separate eSPA models for each lead time of 1, 2, 3, $\ldots$, 24 months, as opposed to a single model that makes predictions for multiple lead times (i.e. multi-horizon prediction). This approach avoids the compounding of model errors at longer lead times, at the expense of having to train multiple models. With eSPA this is a worthwhile trade-off given its excellent scalability properties (being linearly scalable in the number of features $D$, instances $T$, clusters $K$ and classes $M$), which make a single eSPA model very quick to train. This also allows for an easy investigation of the differences in precursors for different lead time predictions through the generation of cluster composites for SST and other fields \cite{Groom2024}. Therefore, for each forecast 50 models are used for each of the 24 lead times, giving a total of 1200 models (although many more models than this are trained during the grid search for each cross-validation split).  

One potential downside to this approach of using separate models to generate independent predictions for each lead time in a continuous forecast is that the predictions at subsequent lead times will not leverage any information about previous predictions that have been made at earlier lead times. Therefore, in addition to the 402 real-valued features, for a lead time of $n$ months we provide as features the class probability distributions up to $n-1$ months. For the training set these will be the true distributions, while for the prediction at $n$ months ahead of the latest available data we provide the mean predictions from the ensemble that have already been made up to $n-1$ months ahead. Thus the predictions by each model at lead time $n$ are conditioned on the sequence of class probabilities that have already been observed/predicted. This is made possible by modifying the clustering metric for categorical features (provided as probability distributions) in eSPA. Rather than use the Euclidean distance as the clustering metric, for categorical features the Kullback-Leibler divergence is used as the appropriate measure and cluster centroids are calculated directly in the probability simplex for each feature, which represents the space of all possible probability distributions over the support of each discrete random variable. The cluster centroid for a categorical feature can be interpreted as the (normalised) geometric mean of the probability distributions assigned to that cluster. For further details, see \ref{app:eSPA}.

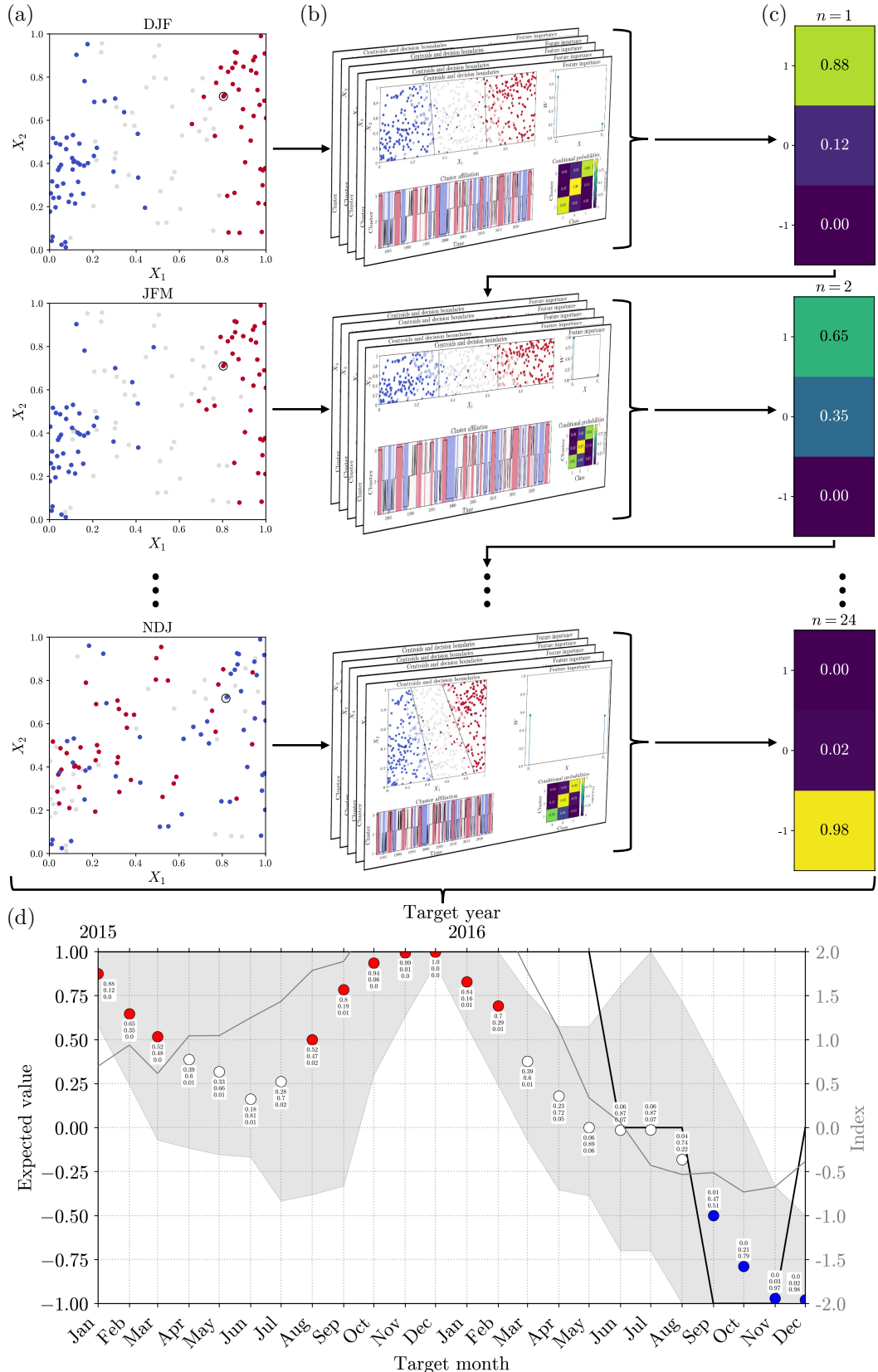
\begin{figure}[H]
  \centering
  \vspace{-3em}
  \begin{tikzpicture}
    \node (img1) {\includegraphics[width=\textwidth]{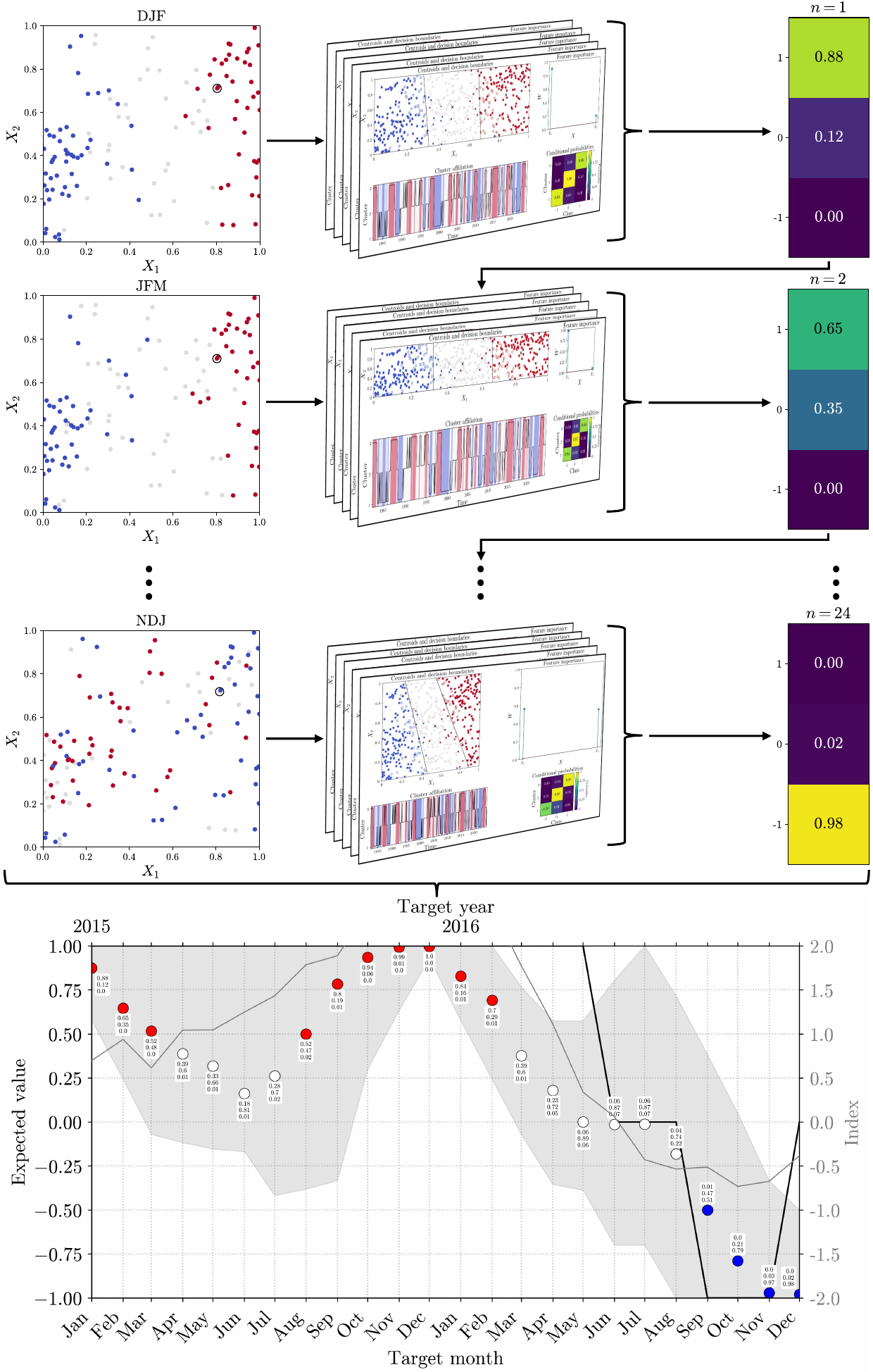}};
    \node[anchor=north west, xshift=0pt, yshift=2pt] at (img1.north west) {(a)};
    \coordinate (b) at ($ (img1.north west)!0.33!(img1.north east) $);
    \node[anchor=north west, xshift=0pt, yshift=2pt] at (b) {(b)};
    \coordinate (c) at ($ (img1.north west)!0.85!(img1.north east) $);
    \node[anchor=north west, xshift=0pt, yshift=2pt] at (c) {(c)};
    \coordinate (d) at ($ (img1.north west)!0.65!(img1.south west) $);
    \node[anchor=north west, xshift=0pt, yshift=0pt] at (d) {(d)};
  \end{tikzpicture}
  \vspace{-2.5em}
  \caption{An illustration of the ensemble learning procedure for generating a 24-month forecast. (a) For each lead time, the data is restricted to only those instances with the same target season. The instance from which the prediction is made is circled. (b) For each lead time an ensemble of eSPA models is fitted to the data. (c) The ensemble-averaged class probabilities are then calculated, with the probabilities at lead time $n$ being used as features for models trained at subsequent lead times. (d) A plot of expected value vs. target month for the forecast. The shading gives the minimum and maximum expected value of the ensemble at each lead time, with the circles denoting the mean.}\label{fig:ensemble-summary}
\end{figure}

Due to the seasonal variability in ENSO precursors, for example due to seasonal footprinting of midlatitude atmospheric variability \cite{Vimont2003} or phase locking of the Indian Ocean Dipole \cite{Saji1999}, it is desirable to have seasonally varying model parameters that cause the model to look for different patterns in the features depending on the target season and lead time \cite{Ham2021}. One straightforward way to achieve this is to train separate models for each target season and lead time \cite{Ham2019,Patil2023}. As noted in \citeA{Ham2021} there are some downsides to this approach, namely that forecast results are generated independently by separate models for each lead time, which can cause the forecast to become less consistent at longer lead times. In the present approach this is handled by the addition of categorical features from previous lead times as described in the paragraph above. Another downside is that by training separate models for each target season the amount of training data is reduced by a factor of 4, which further exacerbates the small data issue. While this may be a limiting factor for other machine learning methods, with eSPA it actually results in both improved generalisation on the validation set (in many cases the best model has a ranked probability score of exactly 0, indicating a perfect fit) as well as more skilful forecasts. This is in spite of the fact that there are now only $\sim 100$ instances available (prior to splitting) for training a given model. A summary of the entropic learning methodology employed for generating forecasts is given in Algorithm 1. Figure \ref{fig:ensemble-summary} provides a visual depiction of the ensemble learning procedure described in Algorithm 1 for a 24-month forecast starting in January 2015.

\begin{algorithm} \label{alg:hindcast}
\caption{Hindcast procedure}
\begin{algorithmic} 
\For{year $y \gets 2012, \dots, 2022$}
    \For{month $m \gets 1, \dots, 12$}
        \State load feature matrix $X$
        \For{lead time $n \gets 1, \dots, 24$}
            \State load class probability matrix $\Pi_n$
            \For{model $i \gets 1, \dots, 50$}
                \State 1. Add categorical features for lead times $0,\ldots,n-1$
                \State 2. Only keep instances with target month $(m+n-1,m+n,m+n+1)$
                \State 3. Split data into 80\% train, 20\% validation
                \State 4. Grid search over hyperparameters $K$, $\varepsilon_E$, $\varepsilon_C$
                \State \hspace{1em} \Return model with lowest RPS on validation set
                \State 5. Make prediction for month $m$
                \State \Return Predicted class probabilities $\hat{\Pi}$
            \EndFor
            \State \Return Average $\hat{\Pi}$ for lead time $n$
        \EndFor
    \EndFor
\EndFor
\end{algorithmic}
\end{algorithm}

The entropic learning methodology described above can also be related to an older format of how forecast information was presented in the IRI ENSO Quick Look from 2002 to 2011 (for example, see \url{https://iri.columbia.edu/our-expertise/climate/forecasts/enso/archive/201112/QuickLook.html}). This older format contained a plot titled "Current Condition vs. Similar Conditions", which displayed the current evolution of the Ni\~no3.4 index over the past 15 months compared with similar evolutions from previous years along with their future trajectories over the following 15 months. This is in essence a simpler version of what eSPA does. Using the previous 15 months of data to define similarity is a form of delay embedding (here we use a 12-month embedding) and the definition of similarity is solely in terms of the conditions in the Ni\~no3.4 region, rather than the entire state of the surface and subsurface ocean - represented through principal components - which is sparsified to isolate the relevant precursors for a given lead time and target season. Aside from this more sophisticated method for determining similarity with previous states of the ocean, the method for calculating probabilities is conceptually the same; given a set of similar conditions at time $t$ (i.e. the set of observations assigned to cluster $k$), use the observed frequencies of each phase at time $t+n$ as the $n$-month ahead prediction. This step is then repeated for each cross-validation split of the training data and each lead time to produce a 24-month ensemble forecast.

\subsection{Post-processing} \label{subec:post-processing}
Rather than use a simple arithmetic average of the model predictions at each lead time, a more advanced aggregation strategy is employed once each model in each hindcast has been trained that takes advantage of the methods for interpreting eSPA models that were demonstrated in \citeA{Groom2024}. The key idea is that, by inspecting various quantities that can be derived from the affiliation matrix $\Gamma$, the cluster centroids $C$, the feature importance vector $W$ and the conditional probability matrix $\Lambda$ of a trained eSPA model (described in detail in \ref{app:eSPA}), that model can be assigned a weight based on how likely it is deemed to be making a correct prediction for the instance corresponding to the start date of the forecast. Rather than perform this assessment manually, it is automated by framing the problem as a binary classification task where, for every single model trained over all of the hindcasts, the features are the various interpretability quantities for that model and the labels are provided by whether the model prediction corresponded to the true ENSO phase in $n$ months time. For the full population of models trained over every hindcast (giving a total of 158,400 models), a separate eSPA model is trained to predict the probability of whether each model made a correct prediction or not. To perform hyperparameter selection while avoiding overfitting, a grid search is performed using 5-fold cross-validation with random shuffling and stratified sampling.

We refer to this new eSPA model as the meta-model, since it borrows ideas from model stacking in machine learning. Crucially however, in our approach the meta-model is not trying to improve on the predictions of the base models in the ensemble, but rather simply assign them a probability based on how likely they are deemed to be making a correct prediction only using features relating to the base models themselves and not the underlying data they were trained on. These probabilities are used to re-weight the average over model predictions at each lead time in each hindcast, which results in an improvement in overall skill. During this re-weighting, three additional heuristics are applied to further filter out bad models from the ensemble:
\begin{enumerate}
    \item Models with a predicted probability less than 0.05 have their weight set to 0.
    \item Models with a predicted probability less than 0.5 have their weight set to 0.
    \item Models predicting El Ni\~no when the mean prediction for that same target month in the previous hindcast was La Ni\~na and vice versa have their weight set to 0.
\end{enumerate}
If at any stage one of these steps results in all models in the ensemble having zero weight, the previous step is reverted to and used to calculate the final weighted average. If step 1 already results in all models having zero weight then the original, unweighted ensemble mean is used as the final prediction. The meta-model probabilities also provide another measure of confidence for the predictions made in a given hindcast, since lead times where none of the probabilities predicted by the meta-model are above a given threshold (e.g. $p=0.05$) can be flagged as being low confidence. A full list of the features used as inputs for training the meta-model is given in \ref{app:meta-model}. Note that all of the results shown in Section \ref{sec:results} have been post-processed using the meta-model aggregation procedure. The same set of results without applying the meta-model are given in Figures S8-S12 of the supporting information.

It should also be noted that, while application of the meta-model constitutes a form of post-processing, the IRI plume results used as the benchmark for comparison in section \ref{sec:results} have also been post-processed as described in the introduction. This post-processing and the methods for calculating probability distributions based on the MME have been directly optimised for the probabilistic skill metrics such as the ranked probability skill score and expected calibration error (which is based on reliability diagrams for each class) that are presented in section \ref{sec:results}, whereas the meta-model has been optimised to classify correct vs. incorrect eSPA models. Therefore, there is scope to further improve the hindcast results through applying similar types of bias and reliability corrections, in addition to the meta-model procedure.

\subsection{Metrics} \label{subsec:metrics}
The following metrics are used both for scoring individual eSPA models as well as for assessing the ensemble predictions against ground truth data. The ranked probability score (RPS) is defined as 
 \begin{equation} \label{eqn:RPS}
    \mathrm{RPS} = \frac{1}{T}\sum_{t=1}^{T}\sum_{m=1}^{M}\left(\sum_{j=1}^{m}\hat\Pi_{j,t}-\sum_{j=1}^{m}\Pi_{j,t}\right)^2,
 \end{equation}
where $\hat\Pi_{m,t}$ and $\Pi_{m,t}$ are the predicted and true probabilities for class $m=1,\ldots,M$ and instance $t=1,\ldots,T$ respectively. The RPS thus penalises predictions that are further away from the ground truth more heavily in cases where the classes are ordinal, with a worst-case value of $M-1$. Similarly, the ranked probability skill score (RPSS) is defined as
 \begin{equation} \label{eqn:RPSS}
    \mathrm{RPSS} = 1-\frac{\mathrm{RPS}}{\mathrm{RPS}_c}
 \end{equation}
where $\mathrm{RPS}_c$ denotes the RPS that is obtained when using climatological probabilities for the predictions \cite{Weigel2007}. By definition, a positive RPSS denotes skill relative to climatology, with a value of 1 denoting perfect skill. 

Another measure that takes into account the ordering of classes is to consider the expected value of the predictions, given by
 \begin{equation} \label{eqn:EV}
    \mathrm{EV_t} = -1\times\hat{\Pi}_{1,t}+0\times\hat{\Pi}_{2,t}+1\times\hat{\Pi}_{3,t}=\hat{\Pi}_{3,t}-\hat{\Pi}_{1,t},
 \end{equation}
and define the predicted class label as 
 \begin{equation} \label{eqn:EV-label}
    \hat{y}_t=\begin{cases}
        1 & \mathrm{if} \quad \mathrm{EV}_t<-1/3 \\
        3 & \mathrm{if} \quad \mathrm{EV}_t>1/3 \\
        2 & \mathrm{otherwise}
    \end{cases}
 \end{equation}
rather than the conventional definition of $\hat{y}_t=\textrm{argmax}(\hat{\Pi}_{:,t})$. This definition penalises predictions that "hedge" by assigning probability mass to both the La Ni\~na and El Ni\~no classes, e.g. a prediction with $\hat{\Pi}_{1,t}=0.4$ and $\hat{\Pi}_{3,t}=0.6$ would have $\mathrm{EV}_t=0.2$ and thus a predicted label of $\hat{y}_t=2$. Given the predicted class label, we then define the accuracy as
 \begin{equation} \label{eqn:acc}
    \mathrm{Accuracy} = \frac{1}{T}\sum_{t=1}^T\mathbbm{1}(\hat{y}_t=y_t),
 \end{equation}
where $\mathbbm{1}$ is an indicator function that evaluates to 1 if true and 0 otherwise. A value of 1 therefore denotes perfect accuracy, while a value of 0 denotes complete inaccuracy. For the problem presented here with 3 classes, randomly guessing the class would give an accuracy of $1/3$ in expectation.

Classifier performance is also assessed through the (macro-averaged) area under the ROC curve (AUC) and expected calibration error (ECE). Here macro-averaging refers to the process of first calculating the AUC/ECE for each individual class in a one vs. rest approach and then averaging the AUCs/ECEs, weighted by their respective class priors, to get a final score. AUC is calculated by numerically integrating the curve of false positive rate vs. true positive rate (the receiver operating characteristic curve) and is bounded between 1 and 0, where a value of 1 denotes perfect classifier performance. A typical reference value for AUC is that of a random classifier, which in expectation has an AUC of 0.5. ECE is calculated as
 \begin{equation} \label{eqn:ECE}
    \mathrm{ECE} = \sum_{n=1}^{N}\frac{|B_n|}{T}|\mathrm{acc}(B_n)-\mathrm{conf}(B_n)|,
 \end{equation}
where the predicted probabilities are divided into $N$ evenly spaced bins $B_n$ of size $|B_n|$ (here $N=5$ bins are used, following \citeA{Tippett2012}) and $\mathrm{acc}(B_n)$ and $\mathrm{conf}(B_n)$ are the accuracy and confidence for each bin, defined as
 \begin{equation} \label{eqn:acc-conf}
    \mathrm{acc}(B_n) = \frac{1}{|B_n|}\sum_{i\in B_n}\mathbbm{1}(\hat{y}_i=y_i), \qquad \mathrm{conf}(B_n)= \frac{1}{|B_n|}\sum_{i\in B_n}\max(\hat{\Pi}_{:,i}),
 \end{equation}
with $\hat{y}_i=\textrm{argmax}(\hat{\Pi}_{:,i})$ and $y_i$ representing the predicted and true labels for instance $i$. ECE can vary between 0 (perfect calibration) and 1 (complete miscalibration). 

Finally, the Wilson score interval\footnotemark\: is used to calculate 95\% confidence intervals on the AUC and Accuracy and bootstrapping is used to calculate 95\% confidence intervals on the RPSS and ECE.

\footnotetext{The Wilson score interval for a proportion $\hat{p}$ is given by $\left(\hat{p}+\frac{z^2}{2n}\pm z\sqrt{\frac{\hat{p}(1-\hat{p})}{n}+\frac{z^2}{4n^2}}\right)/\left(1+\frac{z^2}{n}\right)$ where $n$ is the number of trials and $z$ is the z-score for the desired confidence interval. For Accuracy, $n$ is the total number of predictions $T$, whereas for AUC it is $n_S\times n_F$ where $n_S$ and $n_F$ are the number of successes and failures.}

\section{Results} \label{sec:results}
Using the setup described in Section \ref{sec:methods}, a series of hindcasts are performed. The first hindcast has a start date of January 2012 (i.e. this is the first month to be predicted) and an end date of December 2013 while the last hindcast has a start date of December 2022 and an end date of November 2024. These hindcasts are used to assess the forecast skill according to the metrics detailed in Section \ref{subsec:metrics}, which is compared with the skill of the combined (model-based) probabilistic forecasts produced from the International Research Institute for Climate and Society (IRI) ENSO prediction plume over the same period. 

\begin{figure}[t]
\centering
\begin{tikzpicture}
    \node (img1) {\includegraphics[width=\textwidth]{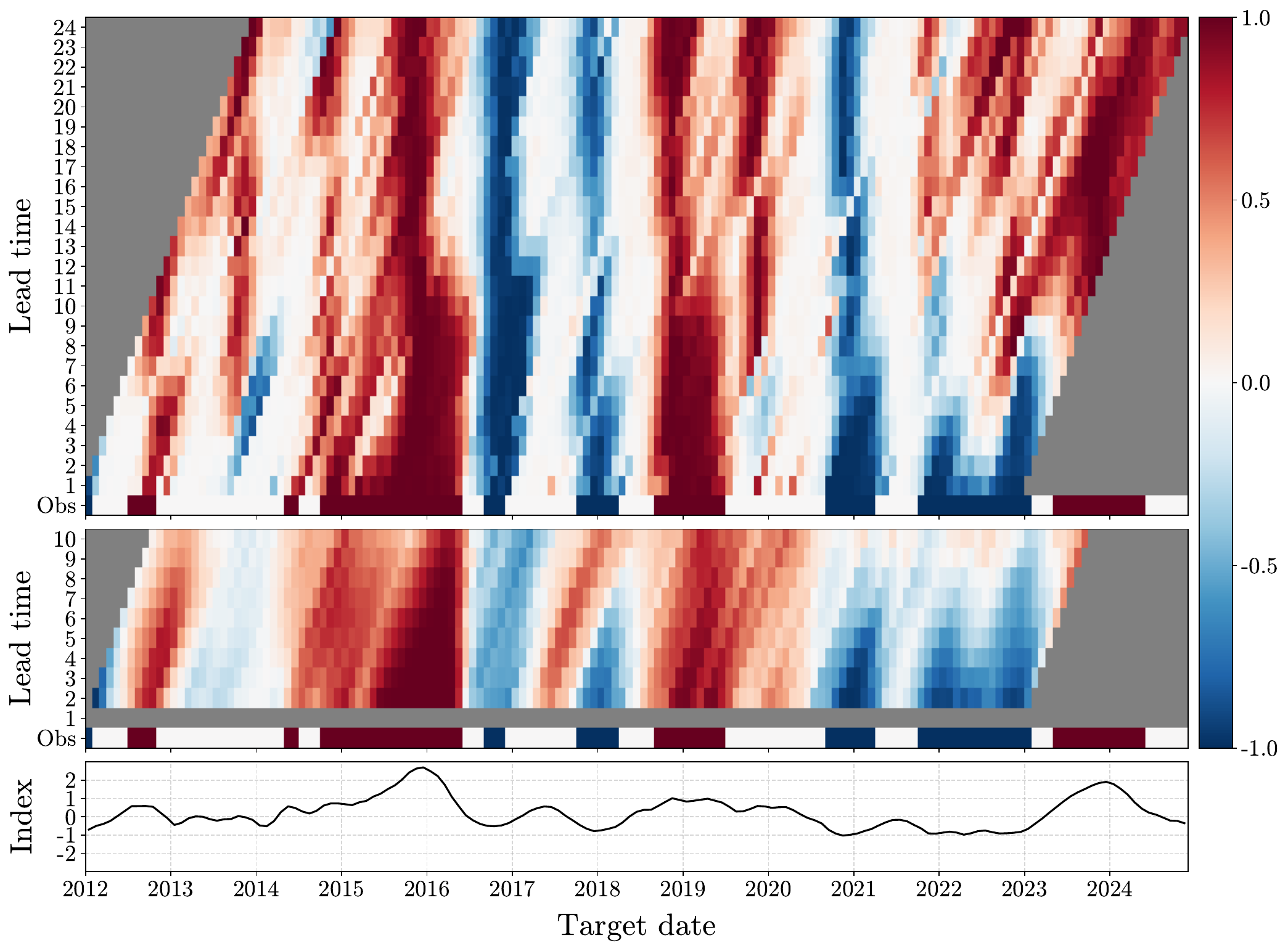}};
    \node[anchor=north west, xshift=28pt, yshift=-8pt] at (img1.north west) {(a)};
    \coordinate (b) at ($ (img1.north west)!0.55!(img1.south west) $);
    \node[anchor=north west, xshift=28pt, yshift=0pt] at (b) {(b)};
    \coordinate (c) at ($ (img1.north west)!0.79!(img1.south west) $);
    \node[anchor=north west, xshift=28pt, yshift=0pt] at (c) {(c)};
  \end{tikzpicture}
\caption{Expected value vs. target date and lead time for (a) eSPA hindcasts and (b) real-time IRI forecasts (b) made over the period 2012-2022. Figure (c) shows the 3-month running average of the Ni\~no3.4 index over this same period.}
\label{fig:hindcasts}
\end{figure}
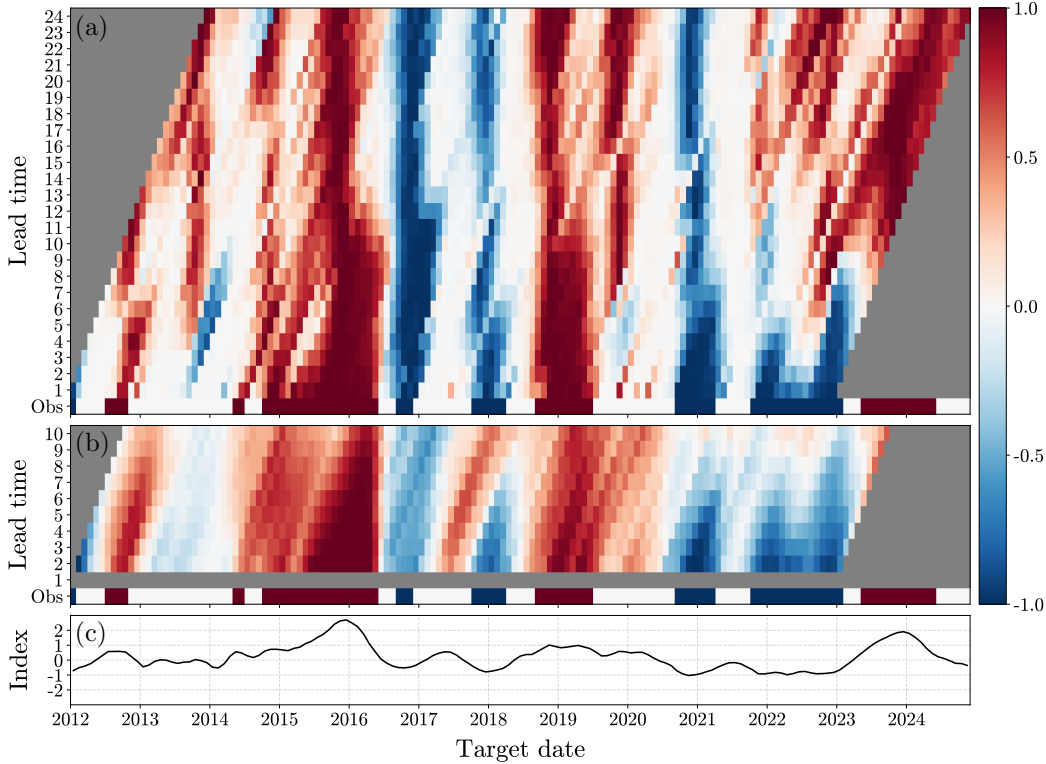

\subsection{Hindcasts}
Figure \ref{fig:hindcasts} plots the results of every hindcast made between January 2012 and December 2022, using a similar convention to Figure 10 of \citeA{Tippett2012}, for both eSPA and the IRI plume. By assigning the El Ni\~no, neutral and La Ni\~na classes a value of 1, 0 and -1 respectively, the predicted probabilities at each lead time and target date are converted to an expected value as per Equation \ref{eqn:EV}. By comparing these expected values with the observed phase of ENSO for a given target date (as determined by the 3-month running average of the Ni\~no3.4 index with a threshold of $\pm0.5^\circ$), a qualitative assessment of forecast skill can be made for each of the main events during this period. In particular, we see that eSPA successfully forecasts the 2015/16 and 2018/19 El Ni\~no events at 24 months lead time as well as the 2016/17, 2017/18 and 2020/21 La Ni\~na events at 24 months lead time. The early period from 2012 to 2014 is less skilfully predicted by both eSPA and the IRI plume, during which the Ni\~no3.4 index remained almost entirely neutral. Similarly, the recent period from 2022 to 2024 which featured the 2nd and 3rd events of the "triple dip" La Ni\~na is also less skilfully predicted, with these events only successfully forecast by eSPA at 12 and 8 months lead time respectively. During the 2019-2020 period an El Ni\~no event is incorrectly predicted by eSPA for lead times of $\ge 6$ months, however inspection of the Ni\~no3.4 index during this period (displayed in the bottom panel of Figure \ref{fig:hindcasts}) shows that it remained close to the threshold of $0.5^\circ$, suggesting that these longer lead forecasts are not unreasonable in their predictions. Similarly, forecasts made by the IRI plume during the 2017-2018 period incorrectly predict an El Ni\~no event during boreal summer, during which the Ni\~no3.4 index came close to the $0.5^\circ$ threshold. However, due to the slippage phenomenon described in \citeA{Barnston2012,Tippett2012}, at longer lead times an El Ni\~no event is predicted by the IRI plume to persist into the 2017/18 boreal winter, when in actual fact a La Ni\~na event occurred. Finally, the most recent 2023/24 El Ni\~no event is shown to be successfully forecast at all lead times considered in this set of hindcasts.

Aside from the skill for individual events, the following general statements can be made regarding the performance of the eSPA-based forecasting system:
\begin{enumerate}
    \item Unlike the categorical forecasts made by the IRI plume, the eSPA results are less affected by "slippage", a phenomenon whereby the predictions are slow to capture the transition into and out of ENSO events, which manifests as a diagonal tilting of the target date vs. lead time plot.
    \item Forecast skill for target dates during the typical peak of ENSO in boreal winter appears to be correlated with the amplitude of a given event.
    \item La Ni\~na events following an El Ni\~no are more skillfully forecast than subsequent La Ni\~na events.
    \item The majority of incorrect predictions are between adjacent classes, i.e. El Ni\~no and neutral or neutral and La Ni\~na. The only period where this observation does not hold is the 2nd and 3rd events of the "triple dip" La Ni\~na. 
\end{enumerate}

In the following subsections, the skill over the hindcast period for both eSPA and the IRI plume will be quantified and stratified according to both lead time and target season. Note that our definition of a lead time of $n$ months (defined as the number of months between the target month and the month the prediction is being made from) corresponds to a lead time of $n-1$ months using the IRI definition (defined as the number of months between the first month of the forecast and the middle month of the targetted 3-month period), hence the 1 month forecasts from the IRI plume are equivalent to our 2 month forecasts and so forth. This adjustment has been made in the figures so that results shown for the same lead time are directly comparable.

\subsubsection{Skill vs. lead time} \label{subsec:lead-time}

\begin{figure}[t]
\centering
\noindent\includegraphics[width=\textwidth]{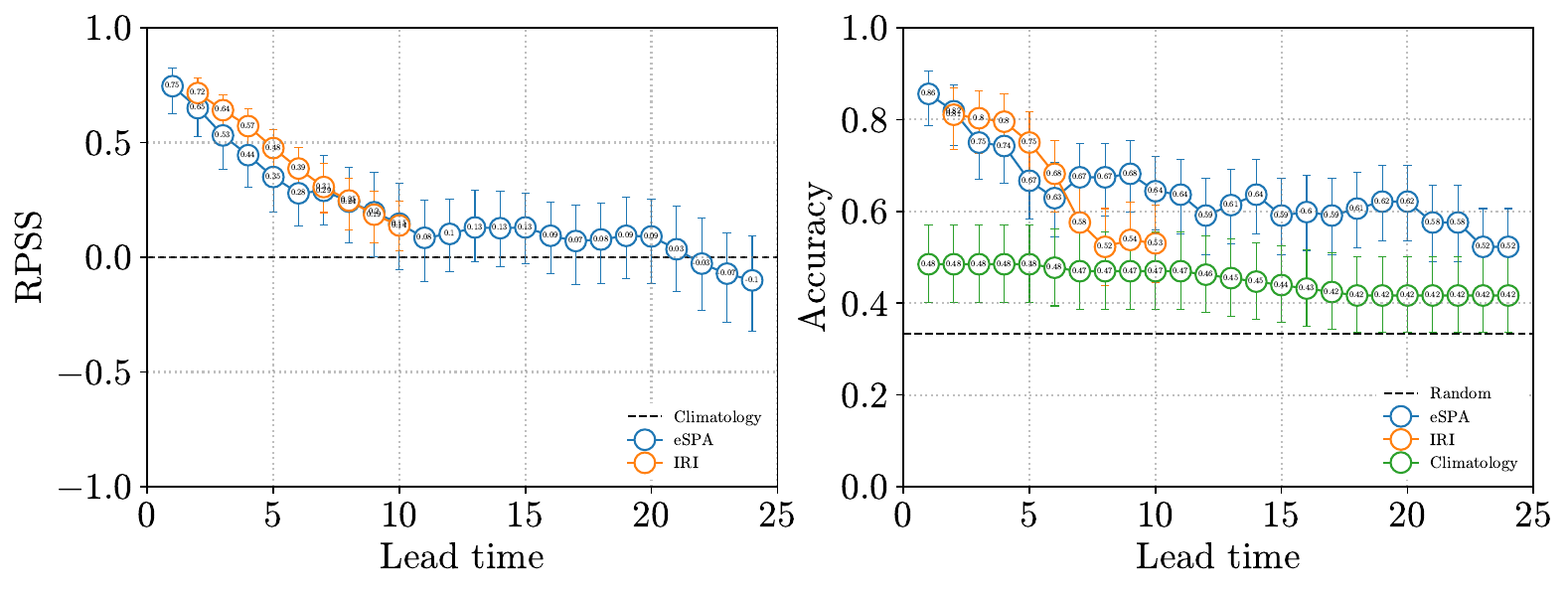}
\caption{Ranked probability skill score and accuracy vs. lead time. Error bars correspond to 95\% confidence intervals, calculated using bootstrapping for RPSS and the Wilson score interval for Accuracy.}
\label{fig:RPSS-lead-time}
\end{figure}

Figure \ref{fig:RPSS-lead-time} shows the ranked probability skill score (higher is better) and accuracy (higher is better) as a function of lead time for forecasts from January 2012 to December 2022 for both eSPA and the IRI plume. In terms of RPSS, eSPA is more skilful than the IRI plume for lead times of 9 months and longer, with skill relative to climatology maintained out to 22 months. Due to the relatively small hindcast period employed in this study, these differences are not statistically significant in terms of 95\% confidence intervals; only the differences between the IRI plume and climatology out to 10 months or eSPA and climatology out to 9 months lead time are statistically significant. In terms of accuracy, eSPA is more skilful than the IRI plume at 2 months lead time as well as lead times of 7 months and longer. As with RPSS, these differences are not statistically significant in terms of 95\% confidence intervals. Compared with predictions based on climatological probabilities, differences with the IRI plume are statistically significant out to 6 months, whereas differences with eSPA are statistically significant for lead times of 1-5 months, 7-10 months, 14 months and 18-20 months. We can therefore conclude that eSPA provides forecasts with similar skill as the IRI plume but at a small fraction of the total computational cost required to run each (dynamical) model that comprises the plume (see \ref{app:cost-estimates} for details on cost estimates), with skill maintained out to more than twice the maximum lead time forecast published by the IRI. Plots of accuracy using the conventional definition rather than our definition based on expected value are given in Figures S1 and S2 of the supporting information and do not alter these conclusions.

\begin{figure}[t]
    \centering
    \begin{subfigure}{0.49\textwidth}
        \centering
        \includegraphics[width=\textwidth,angle=0]{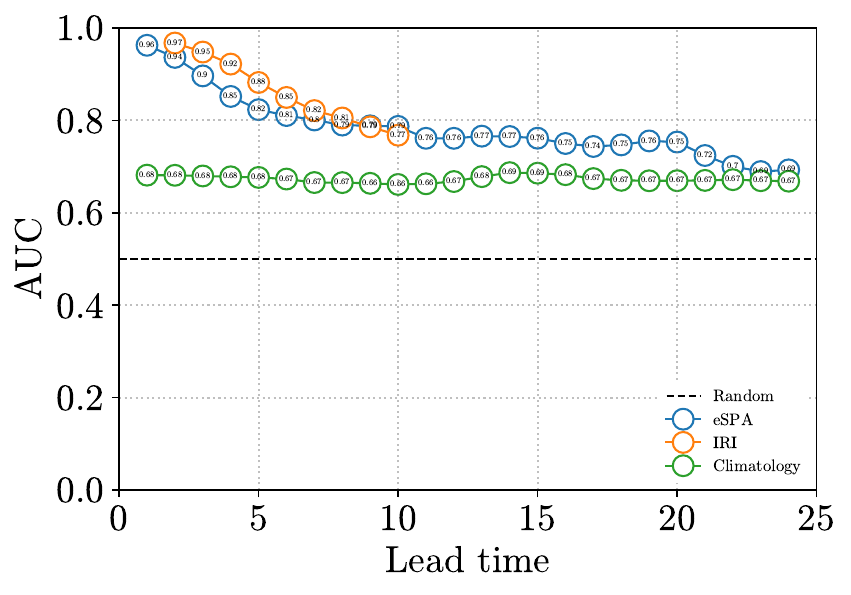}
    \end{subfigure}
    \begin{subfigure}{0.49\textwidth}
        \centering
        \includegraphics[width=\textwidth,angle=0]{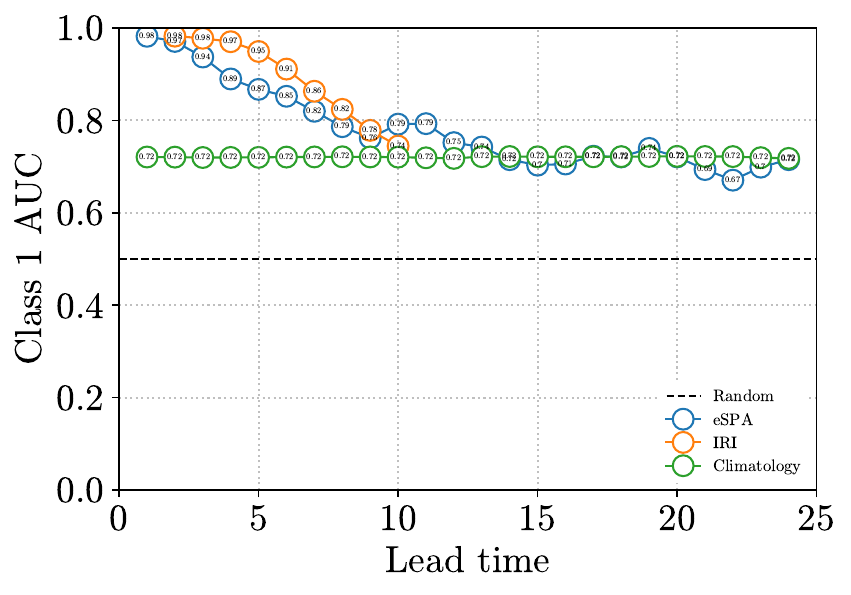}
    \end{subfigure}
    \begin{subfigure}{0.49\textwidth}
        \centering
        \includegraphics[width=\textwidth,angle=0]{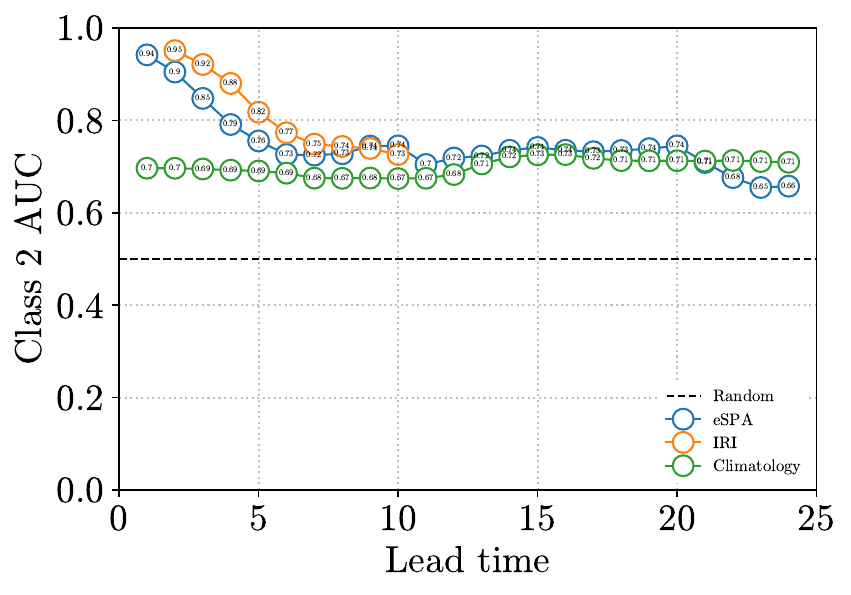}
    \end{subfigure}
    \begin{subfigure}{0.49\textwidth}
        \centering
        \includegraphics[width=\textwidth,angle=0]{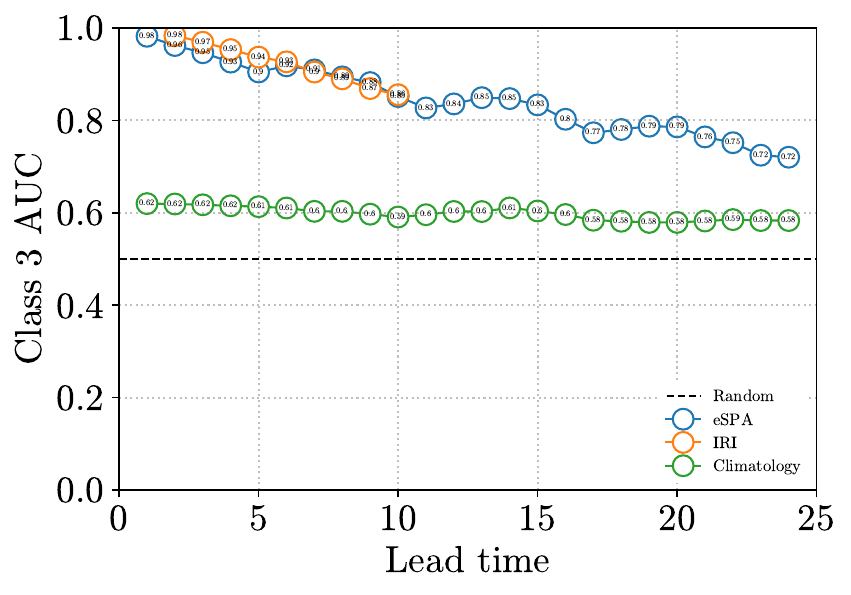}
    \end{subfigure}
    \caption{AUC vs. lead time for the macro-average, class 1 (La Ni\~na), class 2 (Neutral) and class 3 (El Ni\~no). Error bars are plotted corresponding to 95\% confidence intervals but are not visible and are therefore given as tables in \ref{app:confidence-intervals}.}\label{fig:AUC-lead-time}
\end{figure}

\begin{figure}[t]
    \centering
    \begin{subfigure}{0.49\textwidth}
        \centering
        \includegraphics[width=\textwidth,angle=0]{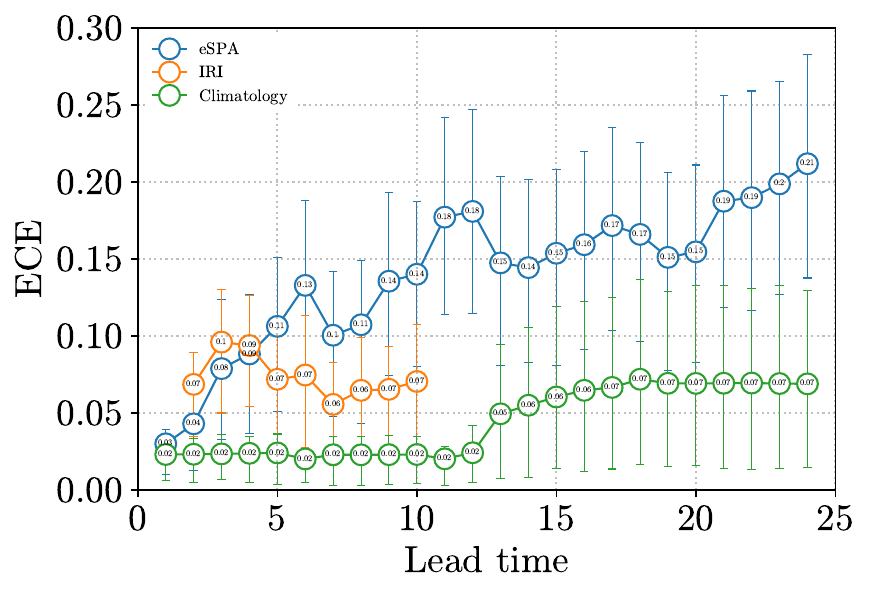}
    \end{subfigure}
    \begin{subfigure}{0.49\textwidth}
        \centering
        \includegraphics[width=\textwidth,angle=0]{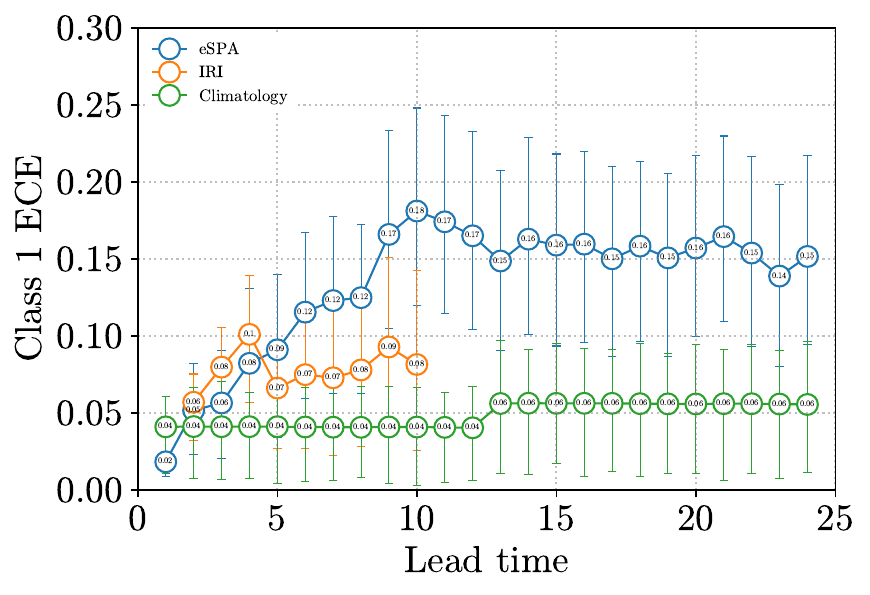}
    \end{subfigure}
    \begin{subfigure}{0.49\textwidth}
        \centering
        \includegraphics[width=\textwidth,angle=0]{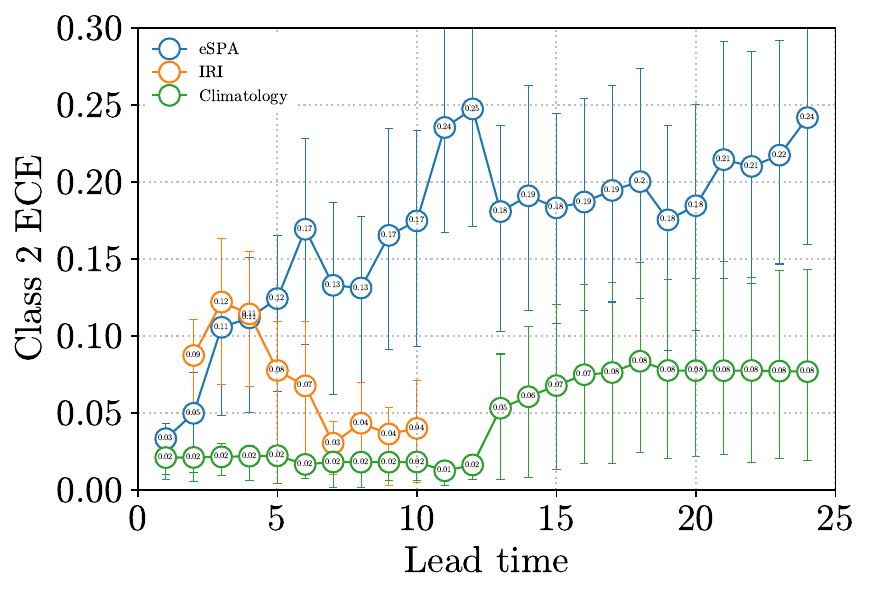}
    \end{subfigure}
    \begin{subfigure}{0.49\textwidth}
        \centering
        \includegraphics[width=\textwidth,angle=0]{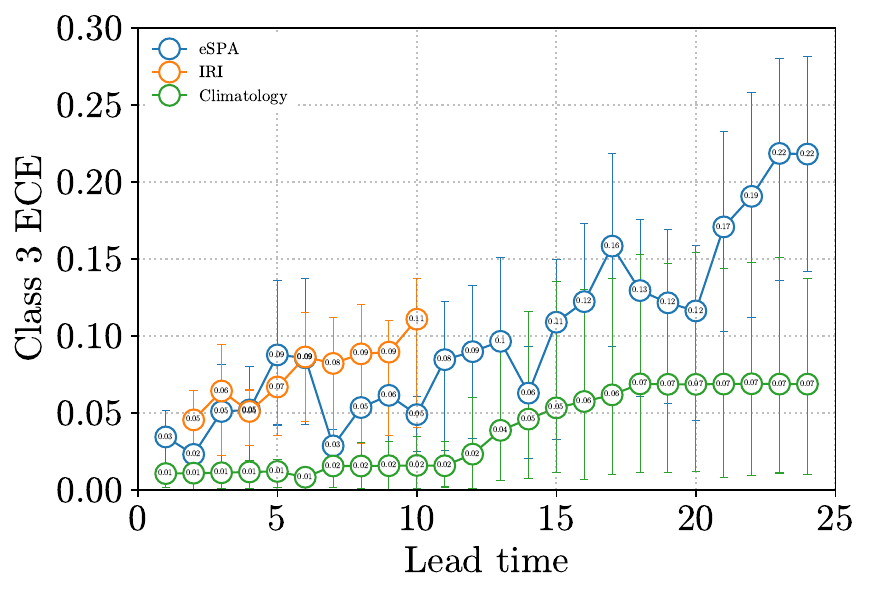}
    \end{subfigure}
    \caption{ECE vs. lead time for the macro-average, class 1 (La Ni\~na), class 2 (Neutral) and class 3 (El Ni\~no). Error bars correspond to 95\% confidence intervals, calculated using bootstrapping.}\label{fig:ECE-lead-time}
\end{figure}

Figures \ref{fig:AUC-lead-time} and \ref{fig:ECE-lead-time} show two alternative metrics that are commonly used to assess classifier performance: the area under the ROC curve (higher is better) and the expected calibration error (lower is better). Both of these metrics are computed as macro-averages of the AUC/ECE for each class, which are plotted in the other subfigures. In terms of AUC, the skill of eSPA is slightly greater than that of the IRI plume for lead times of 9-10 months and remains skilful relative to climatology out to 24 months. Due to the narrower confidence intervals on AUC, the differences between eSPA and the IRI plume are statistically significant for 2-6 months lead time. Differences between the IRI plume and climatological predictions are statistically significant for 2-10 months, while differences between eSPA and climatological predictions are statistically significant for 1-21 months. Note that if a random classifier is used as the skill baseline, as is common in the machine learning literature, rather than climatology then both eSPA and the IRI plume are skilful at the 95\% confidence level for all lead times considered. 

Similar conclusions also hold when looking at the individual class AUCs. For the La Ni\~na class, there are statistically significant differences between eSPA and the IRI plume for 3-8 months, the IRI plume and climatological predictions for 2-9 months and eSPA and climatological predictions for 1-12 months. For the neutral class, there are statistically significant differences between eSPA and the IRI plume for 2-6 months, the IRI plume and climatological predictions for 2-10 months and eSPA and climatological predictions for 1-10 months. Finally, for the El Ni\~no class there are statistically significant differences between eSPA and the IRI plume for 2-5 months, the IRI plume and climatological predictions for 2-10 months and eSPA and climatological predictions for 1-24 months. These differences in class AUCs for eSPA, with predictions for El Ni\~no being more skillful than the other two classes across all lead times, highlight differences in the underlying predictability for each type of event that will be explored in future work. By comparison, the IRI predictions are marginally more skillful for La Ni\~na at lead times of 2-5 months and then El Ni\~no thereafter, suggesting similar underlying mechanisms.

Another useful comparison that can be made is with the state-of-the-art CNN model of \citeA{Patil2023}, who also used the OISST and GODAS datasets for the validation phase (1984 to 2021) of their model. When assessing probabilistic skill, using the same threshold of $\pm0.5^\circ$ to define each class, \citeA{Patil2023} obtained AUCs of 0.69, 0.64 and 0.7 for the El Ni\~no, Neutral and La Ni\~na classes respectively at 24 months lead time. By comparison, eSPA obtains AUCs of 0.72, 0.66 and 0.72 respectively, albeit for a shorter assessment period. In terms of ECE, eSPA is better calibrated at earlier lead times (2-4 months) than the IRI plume and less well-calibrated at longer ones, with none of these differences being statistically significant. Neither eSPA nor the IRI plume is as well calibrated as the climatological probabilities, which is not surprising given that these represent the expected probabilities for each class for a given target season, and for lead times of 3-12 months and 24 months these differences are statistically significant. When looking at individual class ECEs, there are no significant differences between eSPA and the IRI plume for the La Ni\~na class, while for the neutral class the IRI plume is significantly better calibrated for lead times of 7, 9 and 10 months lead time. For the El Ni\~no class, eSPA is better calibrated than the IRI plume for lead times of 2-3 months and 7-10 months, with none of the differences being statistically significant. It is notable that for eSPA, the El Ni\~no class is better calibrated than the other two classes in general, whereas the opposite is true for the IRI plume.

\subsubsection{Skill vs. target season} \label{subsec:target-season}

\begin{figure}[t]
\centering
\begin{subfigure}{\textwidth}
    \centering
    \includegraphics[width=\textwidth,angle=0]{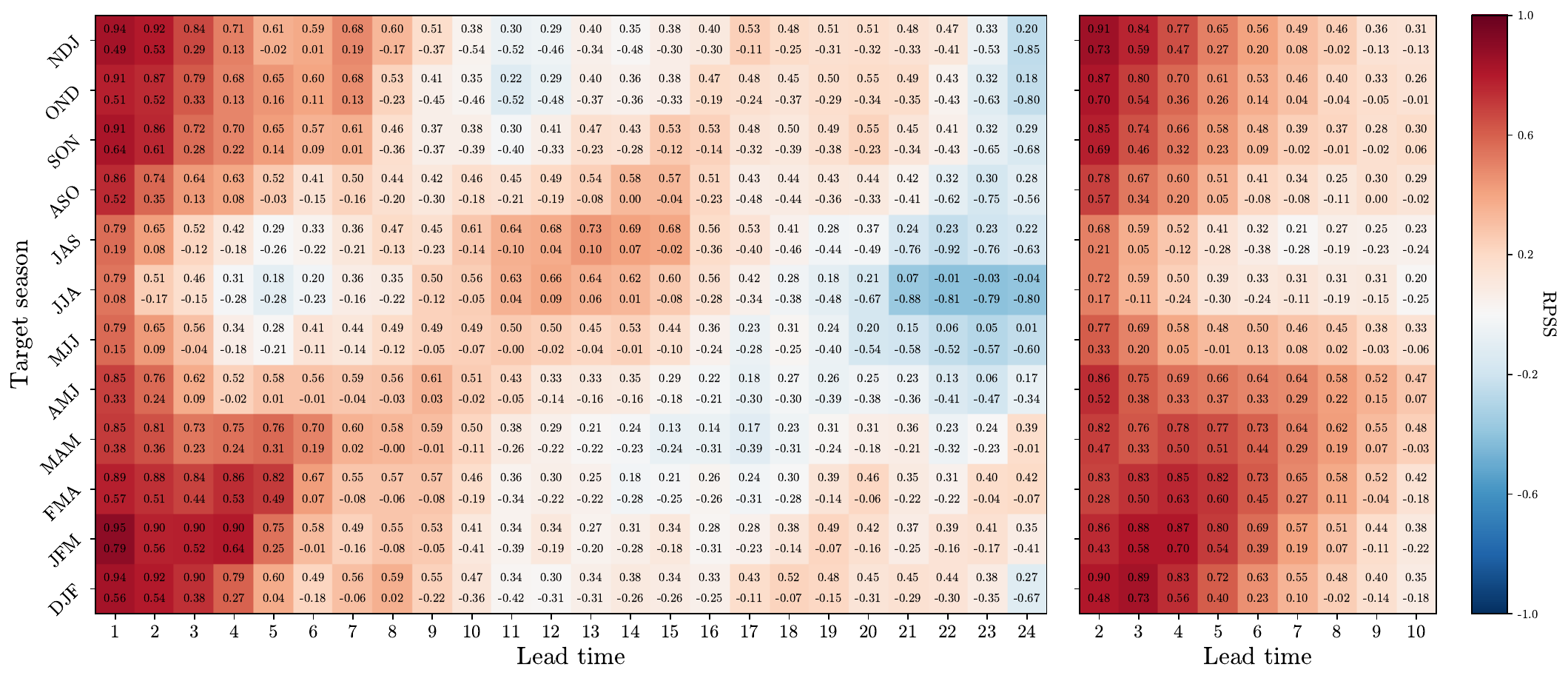}
\end{subfigure}
\begin{subfigure}{\textwidth}
    \centering
    \includegraphics[width=\textwidth,angle=0]{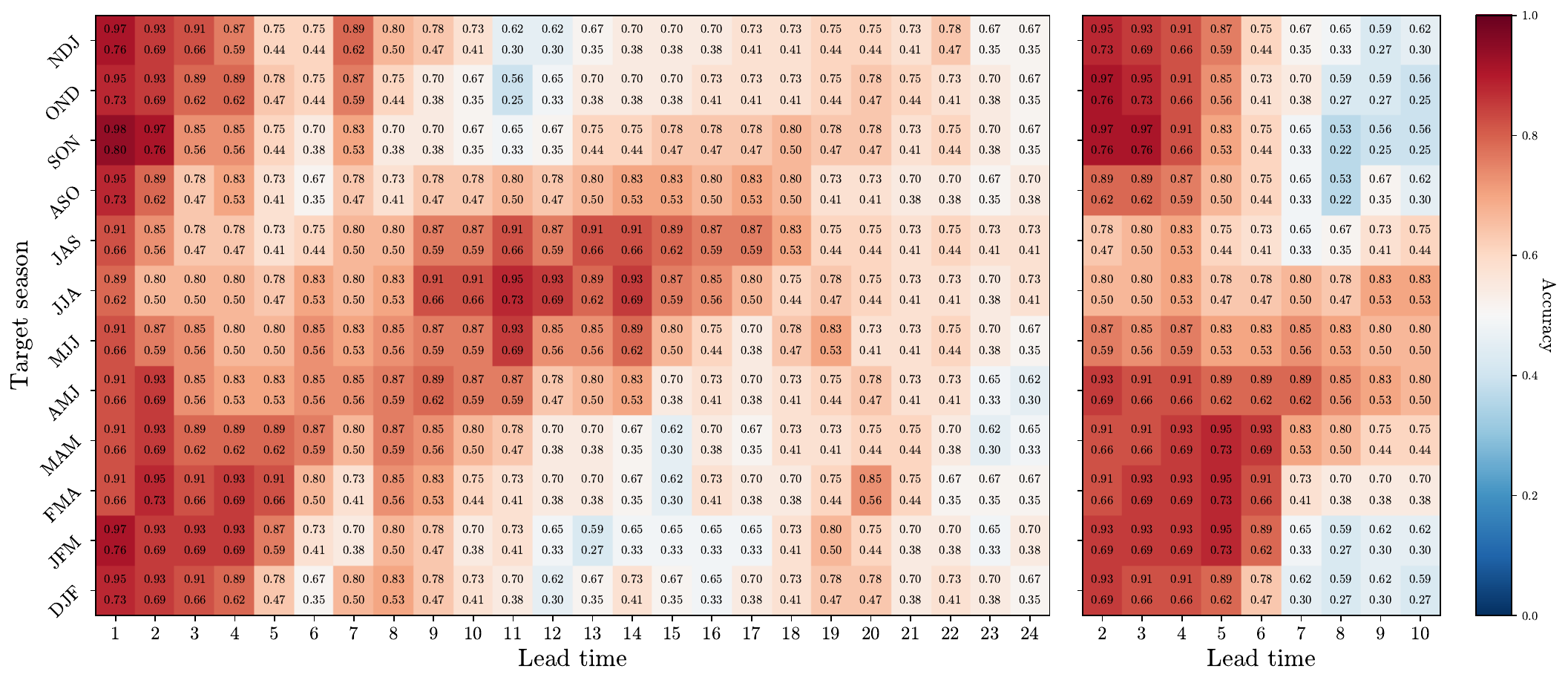}
\end{subfigure}
\caption{Ranked probability skill score and accuracy vs. lead time, stratified by target season for both eSPA (left) and the IRI plume (right). The top and bottom numbers in each cell correspond to the lower and upper bounds of the 95\% confidence interval, calculated using bootstrapping for RPSS and the Wilson score interval for Accuracy.}
\label{fig:RPSS-target-season}
\end{figure}

The skill vs. lead time plots in Figure \ref{fig:RPSS-lead-time} can be further stratified by target season. A caveat to this is that, due to the small hindcast period of 11 years, each target season and lead time combination only contains 33 samples and therefore the results have large confidence intervals associated with them. Nevertheless, Figure \ref{fig:RPSS-target-season} shows the RPSS and Accuracy for both eSPA and the IRI plume as a function of lead time and target season.

In terms of RPSS, there is some evidence of a boreal spring predictability barrier in both the eSPA and IRI results, although any skill barrier that does exist is much less severe than for predictions of the index directly (e.g. see \citeA{Barnston2012}). The target season with the weakest skill for eSPA is JJA, both for shorter lead times of 4-6 months and longer lead times of 18+ months. These regions of low skill are in large part due to misclassifications made for target dates in 2022 during the 2nd and 3rd successive La Ni\~na events, which were incorrectly misclassified as El Ni\~no at those lead times and which can be observed in Figure \ref{fig:hindcasts}. Further examination of Figure \ref{fig:hindcasts} also shows that there are multiple cases where the onset of an event, which typically occurs around JJA, is missed at 5 months lead time. Plots of the (class) AUC(s) vs. lead time and target season are provided in Figures S3-S6 of the supporting information and show a similar dip in skill for JJA at 4-6 months lead, which is mostly due to misclassification made for the La Ni\~na class. Some possible explanations for this are that the use of monthly-averaged data filters out fast-growing modes that are necessary for correct predictions at this lead time, or that the most useful information for prediction is not contained in the provided SST, $\mathrm dT/\mathrm dz$ or $(\tau_x,\tau_y)$ fields at this lead time. 

In terms of the accuracy metric, there is a less significant drop in skill for JJA at 4-6 months lead in the eSPA results. The IRI plume results also show a substantial drop in skill at lead times of 7 months and longer for target seasons of ASO through to JFM, which is not reflected in the eSPA results. This helps to explain where most of the skill advantage for eSPA at these lead times comes from in this metric, which rewards confident predictions of the correct class.  

\section{Conclusions}  \label{sec:conclusion}
This paper has demonstrated the effective application of the entropy-optimal Sparse Probabilistic Approximation (eSPA) algorithm to long-range forecasting of ENSO phase. The eSPA classifier predicts whether the Ni\~no3.4 index will be in El Ni\~no, La Ni\~na, or neutral conditions at a given lead time using a set of features derived from a delay-embedded EOF analysis of global sea surface temperatures, subsurface thermocline proxies and surface wind stresses in the tropical Pacific. In contrast to prior work \cite{Groom2024}, a large ensemble of eSPA models were trained and validated exclusively on observational and reanalysis data from the post-1980 satellite era, with great care taken to avoid any form of information leakage from the future into the training set. A series of hindcast experiments were conducted for start dates from January 2012 to December 2022 at lead times of 1 up to 24 months to assess forecast skill. A state-of-the-art multi-model forecast ensemble – the International Research Institute (IRI) ENSO prediction plume – was employed as a benchmark for skill evaluation. 

A key strength of the entropic learning framework is its interpretability and diagnostic insight. In contrast to black-box deep learning models, eSPA provides transparent probabilistic relationships between observed features and ENSO variability. In this study, these interpretable outputs were exploited to design a novel ensemble aggregation strategy. Rather than weighting all ensemble members equally, the internal structure of each individual eSPA model in the ensemble was used to gauge the likelihood of its prediction being correct and re-weight its contribution to the overall prediction at that lead time. This meta-learning approach effectively learns to "predict the likelihood of the predictions", boosting overall forecast performance and offering a practical example of how explainable machine learning can be harnessed in climate forecasting. 

Despite the limited number of training instances (on the order of only a few hundred monthly samples for each hindcast), eSPA achieved skillful performance across all forecast lead times considered. At lead times overlapping with those published for the IRI plume (up to 10 months), eSPA attained predictive skill with statistically insignificant differences at the 95\% confidence level in terms of both accuracy and the ranked probability skill score. Moreover, at extended lead times beyond the range of the IRI operational forecasts, eSPA maintained positive skill out to 22 months in terms of ranked probability skill score and out to 24 months in terms of accuracy and area under the ROC curve (AUC). This performance effectively doubles the forecast horizon of conventional ENSO outlooks, as the IRI and other operational systems typically issue forecasts only up to one year ahead. Notably, eSPA demonstrated the capability to anticipate major ENSO events during the hindcast period well in advance; the 2015/16 and 2018/19 El Ni\~no events were successfully predicted at 24 months lead time, as were the 2016/17, 2017/18 and 2020/21 La Ni\~na events. Furthermore, these forecasts are achieved at a small fraction of the computational cost required by conventional dynamical models ($\sim$1000-10000$\times$ cheaper; see \ref{app:cost-estimates} for details), underscoring the efficiency of our entropic learning framework for near-term climate prediction. In addition, comparisons with other machine learning-based forecasting methods indicate that the present approach is highly competitive. For instance, the AUC obtained by eSPA at the 24-month lead time exceeds that reported for the recent deep convolutional neural network of \citeA{Patil2023}, which was trained on similar data, highlighting the advantages of the proposed approach even relative to state-of-the-art deep learning models.

Given the promising results presented in this work there are several avenues for future research and development, a few of which are highlighted here. Firstly, there is the potential to adapt the framework to directly predict the Ni\~no3.4 index in a regression setting (e.g. via the SPARTAN algorithm presented in \citeA{Horenko2023}), which may provide additional performance benefits for forecasting ENSO phase due to the ordering of targets being naturally enforced in the problem formulation. Such a system should be evaluated with rigorous hindcast experiments in the same manner as this study. A second useful extension would be to train multiple eSPA/SPARTAN models on different segments or regimes of the historical record, within an adaptive regime-learning framework that is able to handle non-stationarity in the climate system due to interdecadal variability and anthropogenic forcing. This approach could help maintain skill during challenging periods, such as the recent 2022-2023 period, by allowing the forecasts to adjust to varying background conditions. Seasonal variability in predictors could also be handled in a similar manner, e.g. using the temporally-regularised eSPA method presented in \citeA{Bassetti2024}, avoiding the need to restrict the training data for each model to only those instances in the given target season. Finally, the methodology developed here may be applied to other modes of climate variability beyond ENSO. In particular, deploying the entropic learning framework to forecast intraseasonal phenomena such as the Madden–Julian Oscillation, or extending it to multiple outputs for simultaneous prediction of both ENSO and the Indian Ocean Dipole \cite{Ling2022} are important next steps. Pursuing these directions could pave the way toward a unified, data-driven model of intraseasonal to interannual tropical climate variability and improve our understanding of predictability across different timescales as well as interactions between different modes that lead to compound events.

\appendix

\section{The entropy-optimal Sparse Probabilistic Approximation algorithm} \label{app:eSPA}
The entropy-optimal Sparse Probabilistic Approximation (eSPA) algorithm simultaneously performs discretisation of the state space, feature selection and classification by minimising a loss function that contains terms for each of these tasks (i.e. it performs multi-task learning). Here we give a brief introduction to each of these tasks and how they are defined, followed by a presentation of the loss function that eSPA aims to minimise.

Discretisation refers to the notion that, given $T$ observations of the state space $X$, we can assign each observation $X(t)$ into one of $K$ discrete states $S=\{S_1,\ldots,S_K\}$ where $t=1,\ldots,T$ is a data index and $S_k$ is a vector containing the coordinates of discrete state $k$. This assignment is performed according to some measure of similarity (i.e. a distance metric $\mathcal{D}(x,y)$) and is represented by an affiliation vector $\Gamma(t)=\{\Gamma_1(t),\ldots,\Gamma_K(t)\}$ where $\Gamma_k(t)\in[0,1]$ is the probability that $X(t)$ belongs to discrete state $S_k$. In general, the reconstructed state $\hat{X}(t)=S\cdot\Gamma(t)$ will be an approximation to the true state, with the approximation quality being expressed as the sum of all distances between the true and reconstructed states obtained for a particular discretisation $S$. Following \citeA{Gerber2020}, the best possible approximation can be defined as the solution of the following constrained minimisation problem for $\mathcal{L}$ with respect to $S$ and $\Gamma$:
\begin{equation} \label{eqn:SPA}
\mathcal{L}(S,\Gamma)=\frac{1}{T}\sum_{t=1}^T\mathcal{D}\left(X(t),S\cdot\Gamma(t)\right)\rightarrow \min_{S,\Gamma\in\Omega_\Gamma}
\end{equation} 
where the feasible set for $\Gamma$ is given by $\Omega_\Gamma=\{\Gamma_k(t)\in[0,1]\forall k,t:\sum_{k=1}^K\Gamma_k(t)=1\forall t\}$. For the case of Euclidean data, i.e. $X\in\mathbb{R}^{D\times T}$, the Euclidean distance is used and the discretisation consists of a matrix of cluster centroids $C\in\mathbb{R}^{D\times K}$. The loss function then becomes
\begin{equation} \label{eqn:SPA-Euclidean}
\mathcal{L}(C,\Gamma)=\frac{1}{DT}\sum_{t=1}^T\sum_{d=1}^D\left(X_{dt}-\sum_{k=1}^KC_{dk}\Gamma_{kt}\right)^2.
\end{equation} 
Note that the number of clusters $K$ is a hyperparameter that must be set by the user. The constrained minimisation problem given by Equation \ref{eqn:SPA} can be solved via an iterative procedure known as the coordinate-descent method that alternates between finding $S^*$ that minimises $\mathcal{L}(S,\Gamma)$ for fixed $\Gamma$ and finding $\Gamma^*$ that minimises $\mathcal{L}(S,\Gamma)$ for fixed $S$. Theorem 1 in \citeA{Gerber2020} proves that, provided a suitable distance metric $\mathcal{D}(x,y)$ is chosen such that Equation \ref{eqn:SPA} is bounded from below, continuously differentiable and separable in $S$ and $\Gamma$, then the iterations generate a monotonically decreasing sequence of solutions with a computational cost that scales linearly with $D$ and $T$ and $K$. Examples of suitable metrics include the Euclidean distance (given in Equation \ref{eqn:SPA-Euclidean}) and the Kullback-Leibler divergence.

The extension to classification is obtained by considering the following (exact) Bayesian model between two stochastic processes $X(t)$ and $Y(t)$, each with discretisations $S^X=\{S^X_1,\ldots,S^X_K\}$ and $S^Y=\{S^Y_1,\ldots,S^Y_M\}$ and probabilistic representations $\Gamma^X(t)$ and $\Gamma^Y(t)$:
\begin{equation} \label{eqn:SPA-Bayes}
\Gamma^Y(t)=\Lambda\Gamma^X(t)
\end{equation} 
where the matrix $\Lambda\in\mathbb{R}^{M\times K}$ contains the conditional probabilities $\Lambda_{mk}$ that $Y(t)$ is in state $S^Y_m$ if $X(t)$ is in state $S^X_k$. These probabilities are assumed to be stationary, i.e. independent of the data index $t$. For the case where $S^Y=\{S^Y_1,\ldots,S^Y_M\}$ is known and defines a set of discrete classes, then Equation \ref{eqn:SPA-Bayes} provides a way to classify the instances in each discrete state $S^X_k$. Setting $\Gamma^Y(t)=\Pi(t)$ where $\Pi(t)\in\mathbb{R}^M$ is the discrete probability distribution over $X(t)$ belonging to class $m=1,\ldots,M$, then $\hat{\Pi}(t)=\Lambda\cdot\Gamma^X(t)$ is the reconstruction of $\Pi(t)$ for a given discretisation $S^X$. Solving the classification task then becomes a matter of (simultaneously) finding the conditional probabilities $\Lambda$, which are obtained by adding a term to $\mathcal{L}$ that minimises the cross-entropy between $\Pi(t)$ and $\hat{\Pi}(t)$ (equivalent to minimising the Kullback-Leibler divergence $\mathcal{D}_{KL}\left(\Pi(t)||\hat{\Pi}(t)\right)$ when $\Pi(t)$ is constant):
\begin{equation} \label{eqn:SPA-Classifier}
\mathcal{L}(C,\Gamma,\Lambda)=\frac{1}{DT}\sum_{t=1}^T\sum_{d=1}^D\left(X_{dt}-\sum_{k=1}^KC_{dk}\Gamma_{kt}\right)^2-\frac{\varepsilon_C}{T}\sum_{t=1}^T\sum_{m=1}^M\Pi_{mt}\log\left(\sum_{k=1}^K\Lambda_{mk}\Gamma_{kt}\right)
\end{equation} 
where the feasible set for $\Lambda$ is given by $\Omega_\Lambda=\{\Lambda_{mk}\in[0,1]\forall m,k:\sum_{m=1}^M\Lambda_{mk}=1\forall t\}$. Compared to Equation \ref{eqn:SPA-Euclidean}, the additional classification term may be thought of as a term that regularises the clustering problem, with the hyperparameter $\varepsilon_C$ governing the relative importance between these two tasks. The supervised learning paradigm also provides an alternate way to select the number of clusters, compared to what is typically done for standard unsupervised clustering methods, by choosing values for $K$ and $\varepsilon_C$ that maximise the out-of-sample classification performance across different cross-validation splits of the data. Furthermore, due to the choice of metrics the coordinate-descent method can be extended to minimise Equation \ref{eqn:SPA-Classifier}, with the caveat that the cost no longer scales linearly in $K$ or $M$ \cite{Horenko2020}. 

To handle cases where not all features (i.e. dimensions of $X$) are equally important for discretisation and classification, Equation \ref{eqn:SPA-Classifier} can be extended to also perform a third task of feature selection (sparsification) through replacing the average discretisation error over all features $d=1,\ldots,D$ by an expectation with respect to a new vector $W\in\mathbb{R}^D$. $W_d$ represents the probability that feature $d$ contributes to the discretisation error, therefore the feasible set for $W$ is given by $\Omega_W=\{W_{d}\in[0,1]\forall d:\sum_{d=1}^D W_{d}=1\}$. A term is also added to the loss function to maximise the entropy of $W$:
\begin{eqnarray} \label{eqn:eSPA}
\mathcal{L}(C,\Gamma,\Lambda,W) = \frac{1}{T}\sum_{t=1}^T\sum_{d=1}^DW_d\left(X_{dt}-\sum_{k=1}^KC_{dk}\Gamma_{kt}\right)^2+\varepsilon_E\sum_{d=1}^DW_d\log(W_d)\ldots&& \nonumber \\
\ldots- \frac{\varepsilon_C}{T}\sum_{t=1}^T\sum_{m=1}^M\Pi_{mt}\log\left(\sum_{k=1}^K\Lambda_{mk}\Gamma_{kt}\right)&&
\end{eqnarray}
The rationale for maximising the entropy of $W$ is as follows. If $\varepsilon_E=0$ then the optimisation step for $W$ in the coordinate-descent method is a linear programming problem on a simplex of linear constraints. Therefore, in general, the minimum will lie at one of the vertices of the simplex defined by $\Omega_W$. By setting $\varepsilon_E>0$, a convex term is added that causes the minimum to lie inside the boundary of $\Omega_W$, thus regularising the solution for $W$. The choice of this convex term -- given by the entropy of $W$ -- is such that, in the limit of $\varepsilon_E\rightarrow\infty$, $W$ approaches the uniform distribution ($W_d\rightarrow1/D$). This provides the least-biased estimate based on the available information in accordance with the principle of maximum entropy \cite{Jaynes1957a,Jaynes1957b}. $W_d$ can therefore be considered as a measure of the importance of feature $d$. Geometrically, we can think of each feature dimension $d$ as being scaled by $\sqrt{W_d}$, with the discretisation problem being solved in this transformed space. 

Theorem 1 in \citeA{Horenko2020} summarises the monotonicity of convergence to, and regularity of, the optimal solution, which is given by
\begin{equation} \label{eqn:eSPA-solution}
[C^*,\Gamma^*,\Lambda^*,W^*] := \argmin_{\substack{
\Gamma\in\Omega_\Gamma \\
\Lambda\in\Omega_\Lambda \\
W\in\Omega_W
}} \left(\mathcal{L}(C,\Gamma,\Lambda,W)\right).
\end{equation} 
In \citeA{Vecchi2022}, an improved algorithm (referred to as eSPA+) was proposed involving a reordering of the optimisation substeps along with the derivation of closed-form solutions to each of the substeps for the case of a binary discretisation (i.e. $\Gamma_{k,t}\in\{0,1\}\forall k,t$). In this case, by deploying Jensen's inequality, the eSPA loss function can be rewritten as
\begin{equation} \label{eqn:eSPA+}
    \mathcal{L}^+=\frac{1}{T}\sum_{t=1}^T\sum_{d=1}^DW_d\sum_{k=1}^K\Gamma_{kt}\left(X_{dt}-C_{dk}\right)^2+\varepsilon_E\sum_{d=1}^DW_d\log(W_d)-\frac{\varepsilon_C}{T}\sum_{m=1}^M\sum_{t=1}^T\Pi_{mt}\sum_{k=1}^K\Gamma_{kt}\log(\Lambda_{mk}).
\end{equation}
Note that although the closed-form solution for the $W$ substep does not depend on whether the discretisation is binary or fuzzy, closed-form solutions for the $\Gamma$, $C$ and $\Lambda$ substeps may only be obtained for a binary discretisation \cite{Horenko2020,Vecchi2022}. The primary advantage of using a binary discretisation is that now the cost of each of the four substeps scales linearly with $D$, $T$ $M$ and $K$, as proven in Theorem 2 of \citeA{Vecchi2022}. Furthermore, due to Jensen's inequality, the loss function $\mathcal{L}^+$ is an upper bound to the original loss function $\mathcal{L}$. Therefore, even if the optimal $\Gamma$ that minimises $\mathcal{L}$ is fuzzy, minimisation of $\mathcal{L}^+$ will still provide an approximate solution. 

Although the eSPA(+) algorithm converges monotonically, the convergence is only to a local minimum, since both the $\mathcal{L}$ and $\mathcal{L}^+$ loss functions are globally non-convex in general. Multiple random restarts are used to help avoid getting trapped in a local minimum that does not provide good generalisation to unseen data. Training a skilful eSPA(+) model therefore consists of finding a good set of hyperparameters $(K,\varepsilon_E,\varepsilon_C)$ through a grid search, using cross-validation to assess out-of-sample performance and multiple random restarts for each hyperparameter combination to avoid getting trapped in local minima. For brevity, although the eSPA+ algorithm is used in practice, it will simply be referred to as eSPA throughout this paper.

\section{Features used to train the meta-model} \label{app:meta-model}
A given base eSPA model consists of a $K\times T$ affiliation matrix $\Gamma$, a $D\times K$ matrix of cluster centroids $C$, an $M\times K$ matrix of conditional probabilities $\Lambda$ and a $D$-dimensional feature importance vector $W$, as well as an $M\times T^\prime$ array of predictions $\hat{\Pi}$ where $T^\prime$ is the number of unlabelled instances. These matrices/vectors are used to derive the following real-valued features that are supplied as inputs to the meta-model (where $t$ corresponds to the most recent unlabelled instance and $k$ denotes the cluster that instance has been assigned to):
\begin{enumerate}
    \item The difference between the true monthly-averaged Ni\~no3.4 index at time $t$ (i.e. the start date of the forecast) and the Ni\~no3.4 index calculated from the SST anomaly composite corresponding to cluster $k$ (obtained by re-combining the principal components values at the centroid $C(:,k)$ with their respective EOFs \cite{Groom2024}). Note: only the 0-months lag field from the SST composite is considered.
    \item The RMSE between the true monthly-averaged Ni\~no3.4 index from times $t-11,\ldots,t$  and the Ni\~no3.4 index calculated from the SST anomaly composite corresponding to cluster $k$. Note: all fields from the SST composite are considered, corresponding to 12 time snapshots.
    \item The pattern correlation between the true monthly-averaged Ni\~no3.4 index from times $t-11,\ldots,t$  and the Ni\~no3.4 index calculated from the SST anomaly composite corresponding to cluster $k$. Note: all fields from the SST composite are considered, corresponding to 12 time snapshots.
    \item The (area-weighted) RMSE between the true monthly-averaged SST anomaly field at time $t$ and the SST anomaly composite corresponding to cluster $k$, restricted to the tropical Pacific ($20^\circ$S-$20^\circ$N and $120^\circ$E-$80^\circ$W).  Note: only the 0-months lag field from the SST composite is considered.
    \item The  (area-weighted) pattern correlation between the true monthly-averaged SST anomaly field at time $t$ and the SST anomaly composite corresponding to cluster $k$, restricted to the tropical Pacific ($20^\circ$S-$20^\circ$N and $120^\circ$E-$80^\circ$W).  Note: only the 0-months lag field from the SST composite is considered.
    \item The (area-weighted) RMSE between the true monthly-averaged $\mathrm dT/\mathrm dz$ anomaly field at time $t$ and the $\mathrm dT/\mathrm dz$ anomaly composite corresponding to cluster $k$, restricted to the tropical Pacific ($120^\circ$E-$80^\circ$W).  Note: only the 0-months lag field from the $\mathrm dT/\mathrm dz$ composite is considered.
    \item The (area-weighted) pattern correlation between the true monthly-averaged $\mathrm dT/\mathrm dz$ anomaly field at time $t$ and the $\mathrm dT/\mathrm dz$ anomaly composite corresponding to cluster $k$, restricted to the tropical Pacific ($120^\circ$E-$80^\circ$W).  Note: only the 0-months lag field from the $\mathrm dT/\mathrm dz$ composite is considered.
    \item The (area-weighted) RMSE between the true monthly-averaged wind stress anomaly field at time $t$ and the wind stress anomaly composite corresponding to cluster $k$, restricted to the tropical Pacific ($20^\circ$S-$20^\circ$N and $120^\circ$E-$80^\circ$W).  Note: only the 0-months lag field from the wind stress composite is considered.
    \item The  (area-weighted) pattern correlation between the true monthly-averaged wind stress anomaly field at time $t$ and the wind stress anomaly composite corresponding to cluster $k$, restricted to the tropical Pacific ($20^\circ$S-$20^\circ$N and $120^\circ$E-$80^\circ$W).  Note: only the 0-months lag field from the wind stress composite is considered.
    \item A binary variable indicating whether the Ni\~no3.4 index calculated from the SST anomaly composite corresponding to cluster $k$ is in the same phase as the true monthly-averaged Ni\~no3.4 index at time $t$. Note: only the 0-months lag field from the SST composite is considered.
    \item A binary variable indicating whether the composite generated by averaging over the SST anomaly field for all instances that appear $n$ months ahead of those instances assigned to cluster $k$ has a Ni\~no3.4 index that is in the same phase as the majority class of the predicted distribution $\hat{\Pi}(:,t)$ for lead time $n$. 
    \item The distance on the probability simplex between the predicted label $\hat{\Pi}(:,t)$ and the extremised predicted distribution $\tilde{\Pi}(:,t)$, which is calculated as 
    \begin{equation*}
        \tilde{\Pi}(m,t) = \begin{cases}
            1 & \mathrm{if} \quad m = \mathrm{argmax}(\hat{\Pi}(:,t)) \\
            0 & \mathrm{otherwise}
        \end{cases}
    \end{equation*}
    \item The Euclidean distance between the (pre-processed) feature vector $X(:, t)$ and the cluster centroid $C(:, k)$.
    \item The weighted Euclidean distance between the (pre-processed) feature vector $X(:, t)$ and the cluster centroid $C(:, k)$, weighted by $W$.
    \item The minimal adversarial distance \cite{Horenko2023b} from instance $t$ to a cluster $k^\prime$ where $\Lambda(m,k^\prime)\le 1/3$ and $m = \mathrm{argmax}(\hat{\Pi}(:,t))$.
    \item The (two-tailed) $p$-value for cluster $k$ that is calculated by forming a contingency table between $\Gamma_{k,:}$ and $\Pi_{m,:}$ for each $m$ and using Fisher's exact test to calculate the probability of observing this particular arrangement of the data under the null hypothesis that either value of the true probability for class $m$ (i.e. 0 or 1) is likely to be present in the instances assigned to cluster $k$. The $p$-value that is returned for each cluster is the one corresponding to the class with the highest conditional probability (given by $\mathrm{argmax}(\Lambda_{:,k})$).
    \item The fraction of clusters $\tilde{K}$ that have a $p$-value $<0.05$.
    \item The proportion of features $\tilde{D}$ whose feature importance $W_d$ is greater than the maximum entropy limit of $1/D$.
    \item The total weight in $W$ assigned to real-valued features.
    \item The ranked probability score for the training set.
    \item The ranked probability score for the validation set.
    \item The lead time, given as an integer between 1 and 24.
    \item $\cos\left(\frac{\pi}{6}(m-1)\right)$, where m is an integer between 1 and 12 representing the target month.
    \item $\sin\left(\frac{\pi}{6}(m-1)\right)$, where m is an integer between 1 and 12 representing the target month.
\end{enumerate}

These features are pre-processed using a quantile transformation to make them uniformly distributed on the interval $[0,1]$. In addition, the following categorical features are also supplied as inputs to the meta-model:
\begin{enumerate}
    \setcounter{enumi}{24}
    \item The predicted probabilities $\hat{\Pi}(:,t)$.
    \item The predicted probabilities $\hat{\Pi}(:,t-1)$.
    \item The average predicted probabilities for all 50 models with the same start date and lead time $n$.
    \item The average predicted probabilities for all 50 models with the same start date and lead time $n-1$. If $n=1$, this is set to the true class probabilities on the start date.
    \item The average predicted probabilities corresponding to the start of the sequence of all consecutive predictions with $\mathrm{argmax}(\hat{\Pi}(:,t))=m$. This is set to the true class probabilities on the start date if the sequence extends all the way back to $n=0$.
    \item The climatological probabilities for the month corresponding to the target month at time $t + n$.
\end{enumerate}

Figure S7 in the supporting information contains a plot of the probability vector $W$ for the meta-model, highlighting the relative importance of each of the above features. The final meta-model with optimal hyperparameters (chosen using a grid search and 5-fold cross-validation) obtained an AUC of 0.837, indicating that the above list of features provides good insight into whether a given eSPA model is making a correct prediction or not. The two most important real-valued features are feature 12, which can be interpreted as a measure of how confident the model is in its prediction, and feature 11, which when true is an indication that there is an inconsistency between the clustering and estimation of the conditional probabilities in the sparsified feature space vs. if the probabilities were estimated using the same clusters but in the original feature space. The two most important categorical features are features 25 and 26, which when combined provide a measure of the persistence and consistency of the model's predictions. For example, if the prediction at time $t-1$ is a La Ni\~na event but the prediction at time $t$ is an El Ni\~no, this is suggestive that the model has not learned a good representation of the dynamics. For further details on these various interpretability metrics, see \citeA{Groom2024}.

\section{Estimates of computational cost} \label{app:cost-estimates}
In this appendix we compare estimates of the total computational cost, measured in terms of energy usage, to perform a 24-month forecast of ENSO for eSPA vs. a typical seasonal prediction system. These estimates are by no means precise and should only be considered in terms of their relative order of magnitude differences.

The seasonal prediction system considered is the Met Office GloSea5-GC2 system \cite{Williams2015}, which is based on the Global Coupled model 2.0 (GC2). From \citeA{Williams2015}, GC2 is quoted as achieving 1.87 simulated years per wall clock day when run on 36 nodes of an IBM Power7 high-performance computer (each node consisting of four 8-core Power7 chips). Based on a thermal design power (TDP) for each Power7 chip of 240W and assuming that this power is being drawn constantly by each chip then the estimated power consumption of each node is 960W. This is likely an overestimation of the CPU power consumption but neglects all other aspects of the node that also consume power (memory, storage, networking, etc). The total power consumption across 36 nodes is therefore estimated to be 34.56kW. To complete 2 simulated years therefore requires 25.7h of wall clock time and an estimated \textbf{887kWh of energy}. This estimate is for a single ensemble member of the GloSea5-GC2 seasonal prediction system and does not consider any additional factors that would add to the total cost, such as data assimilation or post-processing.

For the entropic learning forecast system detailed here, we start by noting that the average training time for a single eSPA model over the hindcast period was 2.95ms on a single core of an AMD EPYC 7543 processor, which has a TDP of 225W. To compute a 24-month forecast, each month consists of training an eSPA model on 50 separate cross-validation splits of the training data. For each split, a grid search is performed across 512 different hyperparameter combinations and for each hyperparameter combination 32 separate models are fitted, each with different initial guesses. The AMD EPYC 7543 processor contains 32 cores and each initial guess is fitted on a separate core. Using the same assumptions as above regarding power consumption, we arrive at a total of 0.503h wall clock time to complete a 24-month forecast and an estimated \textbf{0.113kWh of energy}. Some of the same caveats as above apply to this estimate, which does not include any additional costs due to post-processing. Nonetheless, based on these estimates we conclude that the full ensemble of 50 eSPA models is between 1000-10000$\times$ cheaper (in terms of energy consumption) to run than a single ensemble member of a state-of-the-art seasonal prediction system.

\section{Confidence intervals for AUC} \label{app:confidence-intervals}
Tables \ref{tab:AUC}-\ref{tab:AUC-3} present the 95\% confidence intervals on the AUC for eSPA, the IRI plume and climatological probabilities that are not easily visible in Figure \ref{fig:AUC-lead-time}. These are calculated using the Wilson score interval for binomial proportions.

\begin{table}[h] 
\caption{95\% confidence intervals for the macro-averaged AUC.}
\centering \label{tab:AUC}
\begin{tabular}{l c c c} 
\hline
Lead time & eSPA & IRI & Climatology \\
\hline
1 & [0.96, 0.97] & - & [0.67, 0.7] \\
2 & [0.93, 0.94] & [0.96, 0.97] & [0.67, 0.7] \\
3 & [0.89, 0.91] & [0.94, 0.95] & [0.66, 0.69] \\
4 & [0.84, 0.86] & [0.91, 0.93] & [0.66, 0.69] \\
5 & [0.81, 0.83] & [0.87, 0.89] & [0.66, 0.69] \\
6 & [0.8, 0.82] & [0.84, 0.86] & [0.66, 0.69] \\
7 & [0.79, 0.81] & [0.81, 0.83] & [0.65, 0.68] \\
8 & [0.78, 0.8] & [0.79, 0.82] & [0.65, 0.68] \\
9 & [0.78, 0.8] & [0.77, 0.8] & [0.65, 0.68] \\
10 & [0.77, 0.8] & [0.75, 0.78] & [0.65, 0.68] \\
11 & [0.75, 0.77] & - & [0.65, 0.68] \\
12 & [0.75, 0.77] & - & [0.65, 0.68] \\
13 & [0.75, 0.78] & - & [0.66, 0.69] \\
14 & [0.75, 0.78] & - & [0.67, 0.7] \\
15 & [0.75, 0.77] & - & [0.67, 0.7] \\
16 & [0.74, 0.76] & - & [0.67, 0.7] \\
17 & [0.73, 0.76] & - & [0.66, 0.69] \\
18 & [0.73, 0.76] & - & [0.66, 0.68] \\
19 & [0.74, 0.77] & - & [0.65, 0.68] \\
20 & [0.74, 0.77] & - & [0.65, 0.68] \\
21 & [0.71, 0.74] & - & [0.66, 0.68] \\
22 & [0.69, 0.71] & - & [0.66, 0.69] \\
23 & [0.67, 0.7] & - & [0.65, 0.68] \\
24 & [0.68, 0.71] & - & [0.65, 0.68] \\
\hline
\end{tabular}
\end{table}

\begin{table}[h] 
\caption{95\% confidence intervals for class 1 (La Ni\~na) AUC.}
\centering \label{tab:AUC-1}
\begin{tabular}{l c c c}
\hline
Lead time & eSPA & IRI & Climatology \\
\hline
1 & [0.98, 0.99] & - & [0.7, 0.74] \\
2 & [0.96, 0.98] & [0.98, 0.99] & [0.7, 0.74] \\
3 & [0.93, 0.95] & [0.97, 0.98] & [0.7, 0.73] \\
4 & [0.88, 0.9] & [0.96, 0.98] & [0.7, 0.74] \\
5 & [0.85, 0.88] & [0.94, 0.96] & [0.7, 0.74] \\
6 & [0.84, 0.86] & [0.9, 0.92] & [0.7, 0.74] \\
7 & [0.81, 0.83] & [0.85, 0.87] & [0.71, 0.74] \\
8 & [0.77, 0.8] & [0.81, 0.84] & [0.71, 0.74] \\
9 & [0.75, 0.78] & [0.76, 0.79] & [0.71, 0.74] \\
10 & [0.78, 0.81] & [0.73, 0.76] & [0.7, 0.74] \\
11 & [0.78, 0.81] & - & [0.7, 0.73] \\
12 & [0.74, 0.77] & - & [0.7, 0.73] \\
13 & [0.73, 0.76] & - & [0.71, 0.74] \\
14 & [0.7, 0.73] & - & [0.71, 0.74] \\
15 & [0.69, 0.72] & - & [0.71, 0.74] \\
16 & [0.69, 0.72] & - & [0.71, 0.74] \\
17 & [0.71, 0.74] & - & [0.71, 0.74] \\
18 & [0.71, 0.74] & - & [0.71, 0.74] \\
19 & [0.72, 0.75] & - & [0.71, 0.74] \\
20 & [0.71, 0.74] & - & [0.71, 0.74] \\
21 & [0.68, 0.71] & - & [0.71, 0.74] \\
22 & [0.65, 0.69] & - & [0.71, 0.74] \\
23 & [0.68, 0.71] & - & [0.7, 0.73] \\
24 & [0.7, 0.73] & - & [0.7, 0.73] \\
\hline
\end{tabular}
\end{table}

\begin{table}[h]  
\caption{95\% confidence intervals for class 2 (Neutral) AUC.}
\centering \label{tab:AUC-2}
\begin{tabular}{l c c c}
\hline
Lead time & eSPA & IRI & Climatology \\
\hline
1 & [0.93, 0.95] & - & [0.68, 0.71] \\
2 & [0.9, 0.91] & [0.94, 0.96] & [0.68, 0.71] \\
3 & [0.84, 0.86] & [0.91, 0.93] & [0.68, 0.71] \\
4 & [0.78, 0.8] & [0.87, 0.89] & [0.68, 0.71] \\
5 & [0.74, 0.77] & [0.81, 0.83] & [0.68, 0.7] \\
6 & [0.71, 0.74] & [0.76, 0.79] & [0.67, 0.7] \\
7 & [0.71, 0.74] & [0.74, 0.76] & [0.66, 0.69] \\
8 & [0.71, 0.74] & [0.73, 0.76] & [0.66, 0.69] \\
9 & [0.73, 0.76] & [0.73, 0.75] & [0.66, 0.69] \\
10 & [0.73, 0.76] & [0.71, 0.74] & [0.66, 0.69] \\
11 & [0.69, 0.72] & - & [0.66, 0.69] \\
12 & [0.7, 0.73] & - & [0.67, 0.7] \\
13 & [0.71, 0.74] & - & [0.69, 0.72] \\
14 & [0.72, 0.75] & - & [0.71, 0.73] \\
15 & [0.73, 0.75] & - & [0.71, 0.74] \\
16 & [0.72, 0.75] & - & [0.71, 0.74] \\
17 & [0.72, 0.75] & - & [0.7, 0.73] \\
18 & [0.72, 0.75] & - & [0.7, 0.73] \\
19 & [0.73, 0.75] & - & [0.7, 0.73] \\
20 & [0.73, 0.76] & - & [0.7, 0.73] \\
21 & [0.69, 0.72] & - & [0.7, 0.73] \\
22 & [0.66, 0.69] & - & [0.7, 0.73] \\
23 & [0.64, 0.67] & - & [0.7, 0.72] \\
24 & [0.64, 0.67] & - & [0.7, 0.72] \\
\hline
\end{tabular}
\end{table}

\begin{table}[h] 
\caption{95\% confidence intervals for class 3 (El Ni\~no) AUC.}
\centering \label{tab:AUC-3}
\begin{tabular}{l c c c}
\hline
Lead time & eSPA & IRI & Climatology \\
\hline
1 & [0.98, 0.99] & - & [0.6, 0.64] \\
2 & [0.96, 0.97] & [0.98, 0.99] & [0.6, 0.64] \\
3 & [0.94, 0.95] & [0.96, 0.97] & [0.6, 0.63] \\
4 & [0.92, 0.93] & [0.95, 0.96] & [0.6, 0.63] \\
5 & [0.89, 0.91] & [0.93, 0.94] & [0.6, 0.63] \\
6 & [0.91, 0.93] & [0.92, 0.93] & [0.59, 0.63] \\
7 & [0.9, 0.92] & [0.89, 0.91] & [0.59, 0.62] \\
8 & [0.88, 0.9] & [0.88, 0.9] & [0.59, 0.62] \\
9 & [0.87, 0.89] & [0.86, 0.88] & [0.58, 0.61] \\
10 & [0.84, 0.86] & [0.84, 0.87] & [0.57, 0.61] \\
11 & [0.81, 0.84] & - & [0.58, 0.61] \\
12 & [0.82, 0.85] & - & [0.59, 0.62] \\
13 & [0.84, 0.86] & - & [0.59, 0.62] \\
14 & [0.84, 0.86] & - & [0.59, 0.63] \\
15 & [0.82, 0.85] & - & [0.59, 0.62] \\
16 & [0.79, 0.81] & - & [0.58, 0.61] \\
17 & [0.76, 0.79] & - & [0.57, 0.6] \\
18 & [0.77, 0.79] & - & [0.57, 0.6] \\
19 & [0.77, 0.8] & - & [0.56, 0.59] \\
20 & [0.77, 0.8] & - & [0.56, 0.59] \\
21 & [0.75, 0.78] & - & [0.57, 0.6] \\
22 & [0.74, 0.77] & - & [0.57, 0.6] \\
23 & [0.71, 0.74] & - & [0.57, 0.6] \\
24 & [0.71, 0.73] & - & [0.57, 0.6] \\
\hline
\end{tabular}
\end{table}

\section*{Open Research Section}
The OISST, ERSSTv5, GODAS and NNR2 datasets are available at the following links: \url{https://downloads.psl.noaa.gov/Datasets/noaa.oisst.v2.highres/}, \url{https://downloads.psl.noaa.gov/Datasets/noaa.ersst.v5/}, \url{https://downloads.psl.noaa.gov/Datasets/godas/} and \url{https://downloads.psl.noaa.gov/Datasets/ncep.reanalysis2/}. Source code for eSPA is available at \url{https://github.com/horenkoi/eSPA}. The supporting information, data and code used to generate the figures are available at \url{https://zenodo.org/records/15111019}.

\bibliography{references}

\end{document}